\documentclass[reqno,11pt]{amsart}
\usepackage{fullpage}
\usepackage{amsfonts}
\usepackage{amssymb}
\usepackage{graphicx}
\usepackage{arydshln}
\usepackage{latexsym}
\usepackage[mathscr]{eucal}
\usepackage{cite}
\usepackage{bm}
\usepackage{mathrsfs}
\usepackage{amsmath,amsthm,accents}
\usepackage[colorlinks,linkcolor=red,linkcolor=blue,linktocpage]{hyperref}
\hypersetup{colorlinks=true, citecolor=blue}
\usepackage[top=1 in,bottom=1 in,left=1.0 in,right=1.0 in]{geometry}
\usepackage{subfigure}
\usepackage{xcolor}
\usepackage{mathtools}
\usepackage{multirow}
\usepackage{caption}

\usepackage{physics}

\allowdisplaybreaks

\numberwithin{equation}{section}

\newtheorem{prop}{Proposition}[section]
\newtheorem{theorem}{Theorem}[section]

\newtheorem{remark}{Remark}[section]

\newcommand{\wh}{\widehat}
\newcommand{\wt}{\widetilde}

\newcommand{\bM}{\boldsymbol{M}}
\newcommand{\bL}{\boldsymbol{L}}
\newcommand{\br}{\boldsymbol{r}}

\newcommand{\ba}{\boldsymbol{a}}
\newcommand{\bC}{\boldsymbol{C}}
\newcommand{\bK}{\boldsymbol{K}}
\newcommand{\bS}{\boldsymbol{S}}

\newcommand{\bX}{\boldsymbol{X}}
\newcommand{\bY}{\boldsymbol{Y}}
\newcommand{\bZ}{\boldsymbol{Z}}
\newcommand{\bu}{\boldsymbol{u}}
\newcommand{\bI}{\boldsymbol{I}}
\newcommand{\bv}{\boldsymbol{v}}
\newcommand{\bw}{\boldsymbol{w}}
\newcommand{\bG}{\boldsymbol{G}}

\newcommand{\be}{\boldsymbol{e}}
\newcommand{\bH}{\boldsymbol{H}}
\newcommand{\bJ}{\boldsymbol{J}}
\newcommand{\bF}{\boldsymbol{F}}

\newcommand{\bs}{\boldsymbol{s}}

\newcommand{\ts}{\,^t \hskip -2pt {\boldsymbol{s}}}

\newcommand{\nn}{\nonumber}
\newcommand{\st}{\hbox{\tiny\it{T}}}

\begin{document}

\title{Symmetric discrete Mikhailov models}	

\author{Xiao-gang Mu,~~Song-lin Zhao$^{*}$ \\
\\\lowercase{\scshape{School of Mathematical Sciences, Zhejiang University of Technology, Hangzhou 310023, P.R. China}}}
\email{*Corresponding Author: songlinzhao@zjut.edu.cn}

\begin{abstract}

The discrete Mikhailov model of \cite{ZMZ2026} lacks conjugation symmetry, which prevents a direct reduction to the discrete Fokas--Lenells (dFL) 
equation and restricts the study of its solution structure. Revisiting the Cauchy matrix scheme, we construct four families of fully conjugation-symmetric 
discrete Mikhailov models by combining the equivalent relations of the auxiliary quantities pairwise. All of these models reduce to the symmetric semi-discrete and continuous Mikhailov models via a two-step continuum limit. The local reduction $\nu=\mu^{*}$ and the nonlocal reduction $\nu=\mathrm{i}\mu_{-}^{*}$
then give four families of local and of nonlocal dFL equations, for which exact Cauchy matrix solution schemes are established: 
the AKNS-type scheme for the local case, and both the AKNS- and KP-type schemes for the nonlocal case. 
We further characterize the dynamics of the obtained solutions in detail.
 
\end{abstract}

\keywords{Discrete Mikhailov model; Discrete Fokas--Lenells equation; Cauchy matrix scheme; Reduction; Solutions.}

\maketitle

\section{Introduction}
\label{sec-1}

The Mikhailov model is a relativistically invariant system in $(1+1)$ dimensions, 
originating from studies of the classical massive Thirring model (MTM) and nonlinear equations associated with quadratic spectral problems \cite{GIK1980}. 
Under a standard normalization, it admits the coupled form
\begin{subequations}
\label{MM}
\begin{align}
& u_{xt}+u-2\mathrm{i}uvu_x=0, \\
& v_{xt}+v+2\mathrm{i}uvv_x=0,
\end{align}
\end{subequations}
where $u(x,t)$ and $v(x,t)$ denote complex-valued fields, and $\mathrm{i}$ stands for the imaginary unit.
Upon imposing the complex-conjugation reduction $v=u^{*}$, the coupled system reduces to
\begin{align}
\label{cFL}
u_{xt}+u-2\mathrm{i}|u|^{2}u_x=0,
\end{align}
which can be transformed to the Fokas--Lenells (FL) equation --- a novel generalization of the nonlinear Schr\"{o}dinger (NLS) equation \cite{LF2009}.
The FL equation was first derived using bi-Hamiltonian structures of the NLS equation \cite{F1995}
and governs the propagation of nonlinear pulses in single-mode optical fibers incorporating specific higher-order 
nonlinear effects \cite{L2009}. We also refer to \eqref{cFL} as the FL equation, with no further distinction.
This relation makes the Mikhailov model a natural starting point for studying FL equation through symmetry reductions.
It has been established that the Lax pair of the MTM in light-cone coordinates is gauge equivalent to the Kaup–Newell (KN) spectral problem, 
and that solutions to the MTM can be constructed by solving the FL equation \eqref{cFL} \cite{GIK1980,KN1977}; see also Appendix A in \cite{LWZ2022}.
From the perspective of integrable hierarchies, the Mikhailov model corresponds to the first negative flow of the KN hierarchy in potential variables, 
while the derivative NLS equation arises among its positive flows \cite{CY2008,FGZ2013}. 
Besides the complex-conjugation reduction $v=u^{*}$ to yield the FL equation \eqref{cFL}, the Mikhailov model \eqref{MM} also admits two
so-called nonlocal reductions: $v=u_{-}$\footnote{For the function $f:=f(x_1,x_2)$, notation $f_{-}$ means $f_{-}=f(-x_1,-x_2)$.} 
and $v=\mathrm{i}u^*_{-}$. As a result, one has the first type nonlocal FL equation
\begin{align}
\label{nFL-1}
 u_{xt}+u-2\mathrm{i}uu_{-}u_x=0, 
\end{align}
and the second type nonlocal FL equation
\begin{align}
\label{nFL-2}
u_{xt}+u+2uu^*_{-}u_x=0.
\end{align}
Note that nonlocal integrable systems were first systematically proposed by Ablowitz and Musslimani in 2013 in \cite{AM2013} and have received intensive
attention (e.g., Refs. \cite{F2016,AM2016,AM2017,YY2018,L2018,Z2018,CDLZ2018,GPZ2020,RCPMH2020,RS2021,XZS2023,WLWZ2026,FZF2026}).

A variety of analytical methods have been developed for the Mikhailov model \eqref{MM} and FL type equations \eqref{cFL}-\eqref{nFL-2}. These include  
direct linearization approach \cite{FCQL1983}, inverse scattering transforms and Riemann--Hilbert problems \cite{LF2009,AX2019,ZF2021}, dressing chain \cite{L2010}, 
algebro-geometric method \cite{ZF2013}, Darboux transformations \cite{HXP2012,XHCP2015,ZZY2019,WXL2020}, variable separation technique \cite{WHQGM2019},
bilinear approach \cite{LWZ2022,SLA2023} and Kadomtsev–Petviashvili (KP) reduction scheme \cite{YC2023,ZMF2024}.
It should be stressed that the Mikhailov model \eqref{MM} and the FL equation \eqref{cFL} admit solutions in Cauchy matrix form. 
A chain of B\"{a}cklund transformations of the Mikhailov model \eqref{MM} \cite{V2011} was constructed and viewed as semi-discrete equations
in the Merola--Ragnisco--Tu hierarchy (cf. \cite{MRT1994}) and solutions of the FL equation \eqref{cFL} were
given in terms of Cauchy matrix by using the connection between the Merola--Ragnisco--Tu hierarchy
and the Ablowitz–Ladik hierarchy. Subsequently, 
Matsuno performed a purely algebraic procedure using an elementary theory of determinants 
to show the bright soliton solutions and dark soliton solutions 
to the FL equation \eqref{cFL}, where the $\tau$-function was expressed as Cauchy matrix determinant \cite{Mat2012-1,Mat2012-2}.
Very recently, a reduction from the self-dual Yang-Mills equation to the Mikhailov model 
\eqref{MM} was described in \cite{LLZ2025}, where two types Cauchy matrix solutions of the KP
hierarchy and the AKNS hierarchy were revealed for the Mikhailov model \eqref{MM} and FL equation \eqref{cFL}.
Moreover, the bidifferential calculus approach has also been applied in \cite{MY2025} to construct solutions of the Mikhailov model, \
and the solutions obtained are of Cauchy-matrix type.

The Cauchy matrix approach is an algebraic direct method for integrable systems, originating from the work of Nijhoff, Atkinson and Hietarinta \cite{NAH2009} on 
constructing soliton solutions for the Adler--Bobenko--Suris lattice equations \cite{ABS2003}, where the core algebraic constraint is the Sylvester matrix equation 
governing a dressed Cauchy matrix together with auxiliary vectors satisfying linear dispersion relations. 
This scheme is strictly connected with the direct linearization method
proposed by Fokas and Ablowitz in \cite{FA1981} and hence being a byproduct of the latter one.
Distinct from the inverse scattering transform and Darboux transformation, which usually start from a given Lax pair, the Cauchy matrix approach is 
capable of simultaneously deriving integrable equations and their exact solutions within a unified algebraic framework \cite{ZZ2013}. 
In \cite{XZZ2014}, Xu et al. adopted the Cauchy matrix approach to reveal the links between the Sylvester equation and some continuous integrable systems, 
where the Korteweg--de Vries (KdV), modified KdV, Schwarzian KdV and sine-Gordon (sG) equations were 
taken as examples. Since then, this method has proved effective in studying continuous and 
semi-discrete integrable equations (e.g., literature \cite{ZSL2018,LQYZ2022,FZ2022,LQZ2023,SLZZZ2025,HLZ2026,HZ2026,TLLLY2026}).  

Beyond its use in the continuous and semi-discrete settings, the more important role of the Cauchy matrix approach lies in revealing 
the underlying structures of discrete integrable systems \cite{HJN2016}, including the construction of integrable discretizations of nonlinear partial 
differential equations (see, e.g., \cite{NA2010,ZZS2012,Z2016,Z2019,HSZ2025}) and the derivation of integrability 
attributes of discrete equations such as Lax pairs \cite{CWZ2023}. 
Taking advantage of this feature, in \cite{ZMZ2026} we adopt the Cauchy matrix scheme to study an integrable discretization of the Mikhailov model, in which a pair 
of discrete Miura transformations is constructed. Nevertheless, the discrete Mikhailov model thus obtained has a shortcoming, in that the resulting systems 
do not admit a one-field reduction; accordingly, no discrete versions of the local and nonlocal FL type equations have yet been formulated within the Cauchy matrix framework.

Motivated by these open issues, we revisit the discrete Mikhailov model within the Cauchy matrix framework, with the aim of recovering its conjugation symmetry 
and establishing both local and nonlocal dFL equations. To this end, we observe that the combinations involving $u_{1}$
in the matrix compatibility conditions admit two equivalent representations. 
By consistently pairing the alternative representations in the two compatibility equations, we construct four families of conjugation-symmetric discrete 
Mikhailov models, and show that the asymmetric model reported in \cite{ZMZ2026} arises from cross-pairing the equations associated with two different supplementary 
equivalent forms (see Appendix A). Their continuum-limit consistency of the discrete models is verified by a two-step straight continuum limit: the first step reduces the 
discrete models to symmetric
semi-discrete Mikhailov models that are discrete in time and continuous in space, and the second step recovers the continuous Mikhailov model.
Building on the symmetric discrete Mikhailov models, we next carry out two classes of symmetry reductions. The local reduction $\nu=\mu^{*}$ yields four families of 
local dFL equations, while the nonlocal constraint $\nu=\mathrm{i}\mu_{-}^{*}$
imposed after recombining the lattice equations pairwise yields four families of nonlocal dFL equations. We establish an AKNS-type scheme for 
the local equations and both AKNS- and KP-type schemes for the nonlocal equations.
For all these dFL equations we establish a unified exact-solution scheme within the Cauchy matrix approach and obtain 
explicit soliton and multiple-pole solutions. A detailed dynamical analysis is also presented: for local solitons, we 
give analytic formulas for the peak amplitude, the propagation velocity and the peak trajectory on the \((n,m)\) lattice, 
and rigorously establish that two-soliton collisions are elastic, with phase shifts that depend explicitly on the spectral parameters. 
We also derive explicit conditions for denominator zeros in the nonlocal one-soliton solutions.

The rest of the paper is organized as follows. Section 2 reviews the fundamentals of the Cauchy matrix approach and key master function relations,
then constructs four families of conjugate-symmetric discrete Mikhailov models and explains the origin of asymmetry in the existing formulation. 
Section 3 outlines the AKNS-type and KP-type Cauchy matrix solution schemes for the symmetric discrete Mikhailov models, covering both diagonal-spectrum 
soliton solutions and Jordan-block multiple-pole solutions. Section 4 carries out the two-step continuum limit procedure and derives the corresponding 
semi-discrete and continuous Mikhailov models. Section 5 implements local and nonlocal symmetry reductions, 
presents exact solutions for the resulting local and nonlocal dFL equations, and analyzes their typical dynamical behaviors. Concluding remarks and perspectives for future 
research are summarized in Section 6. For completeness, three appendices are provided at the end of the paper: Appendix A presents two supplementary equivalent forms 
of the discrete Mikhailov model and elaborates on the structural origin of the asymmetry in the previously reported discrete formulation; Appendix B collects
the remaining three families of local dFL equations obtained by applying the local reduction to the other symmetric Mikhailov models; 
Appendix C lists the remaining three families of nonlocal dFL equations derived from the pairwise-recombined lattice systems.

\section{Symmetric discrete Mikhailov model}\label{sec-2}

In this section, we derive four families of symmetric discrete Mikhailov models to address the asymmetry problem of the discrete Mikhailov model constructed in Ref. \cite{ZMZ2026}. Our derivation is carried out within the general framework of the Cauchy matrix approach developed in \cite{ZMZ2026}. Instead of reproducing the complete derivation of the fundamental relations, we concentrate on the modified procedures that yield symmetric formulations.

\subsection{Review of the Cauchy matrix method and key relations}
\label{sec-2-1}

We begin with the determining equation set (DES), which comprises a Sylvester equation together with discrete dispersion relations:
\begin{subequations}
\label{DES}
\begin{align}
\label{SE1}
&\bK\bM-\bM\bL=\br\bs^{\st},\\
\label{sh-r}
& p\wt{\br}=p\br+\bK\br\ba,\quad q\wh{\br}=q\br+\bK^{-1}\br\ba,\\
\label{sh-s}
& p\wt{\bs}+\bL^{\st}~\wt{\bs}\ba=p\bs,\quad q\wh{\bs}+(\bL^{-1})^{\st}~\wh{\bs}\ba=q\bs,
\end{align}
\end{subequations}
where $\bK,\bL$ are invertible constant matrices in $\mathbb{C}^{N\times N}$,
$\bM\in \mathbb{C}^{N\times N}[n,m]$
and $\br, \bs \in \mathbb{C}^{N\times 2}[n,m]$ are matrix functions of $(n,m)\in \mathbb{Z}^2$.
Here $\ba=\mathrm{diag}(1,-1)$ denotes the third Pauli matrix $\sigma_{3}$, while $p$ and $q$ are nonvanishing real lattice spacing parameters along the $n$-direction and $m$-direction, respectively.
The tilde $\wt{\phantom{a}}$ and hat $\wh{\phantom{a}}$ denote forward shifts in $n$ and $m$, namely $\wt{f}=f(n+1,m),~\wh{f}=f(n,m+1)$, and $\wh{\wt{f}}=f(n+1,m+1)$.

From the DES \eqref{DES}, we introduce a family of master functions
\begin{equation}
	\label{Sij}
	\bS^{(i,j)}=\bs^{\st}\bL^{j}(\bC+\bM)^{-1}\bK^{i}\br, ~~~ i,j\in\mathbb{Z}, 
\end{equation}
where $\bC\in\mathbb{C}^{N\times N}$ is chosen such that $\bC+\bM$ is invertible. Under the commutation constraint
\begin{align}\label{kc-cl}
	\bK\bC-\bC\bL=\bm 0,
\end{align}
the master functions $\bS^{(i,j)}$ satisfy the following recurrence relations and shift relations:
\begin{itemize}
\item
Recurrence relations:
\begin{subequations}
\label{Sij-re=1}
\begin{align}
\label{Sij-re=1-po}
\bS^{(i,j+1)}=&\bS^{(i+1,j)}-\bS^{(0,j)}\bS^{(i,0)},\\
\label{Sij-re=1-ne}
\bS^{(i,j-1)}=&\bS^{(i-1,j)}+\bS^{(-1,j)}\bS^{(i,-1)}.
\end{align}
\end{subequations}
\item
Shift relations along the $n$- and $m$-directions:
\begin{subequations}
\label{sh-Sij}
\begin{align}
\label{p-Sij-1}
p\wt{\bS}^{(i,j)}+\ba\wt{\bS}^{(i,j+1)}
=&p\bS^{(i,j)}+\bS^{(i+1,j)}\ba-\bS^{(0,j)}\ba\wt{\bS}^{(i,0)},\\
\label{p-Sij-2}	
p\bS^{(i,j)}-\ba\bS^{(i,j+1)}=&p\wt{\bS}^{(i,j)}-\wt{\bS}^{(i+1,j)}\ba+\wt{\bS}^{(0,j)}\ba\bS^{(i,0)},\\
\label{q-Sij-1}
q\wh{\bS}^{(i,j)}+\ba\wh{\bS}^{(i,j-1)}
=&q\bS^{(i,j)}+\bS^{(i-1,j)}\ba+\bS^{(-1,j)}\ba\wh{\bS}^{(i,-1)},\\
\label{q-Sij-2}
q\bS^{(i,j)}-\ba\bS^{(i,j-1)}=&q\wh{\bS}^{(i,j)}-\wh{\bS}^{(i-1,j)}\ba-\wh{\bS}^{(-1,j)}\ba\bS^{(i,-1)}.
\end{align}
\end{subequations}
\end{itemize}

\begin{remark}\label{Rm-2-1}
The constraint (2.3) gives rise to two canonical Cauchy matrix schemes:
\begin{itemize}
\item
\textbf{AKNS-type}: For $\bL=\bK$, we restrict to invertible matrices $\bC$ commuting
with $\bK$, which can be normalized to the identity matrix $\bI$ by a corresponding transformation of the matrix data.
\item
\textbf{KP-type}: For $\bK$ and $\bL$ sharing no common eigenvalues, the Sylvester Theorem\footnote{
For the known matrices $\bX, \bY$ and $\bZ$, the Sylvester equation
\begin{equation}
\bX\bM-\bM\bY=\bZ
\end{equation}
has a unique solution $\bM$ if and only if
$\mathcal{E}(\bX)\bigcap \mathcal{E}(\bY)=\varnothing$, where $\mathcal{E}(\bX)$ and $\mathcal{E}(\bY)$ denote the spectra of the matrices $\bX$
and $\bY$, respectively.} enforces $\bC=\bm 0$.
\end{itemize}
The explicit forms of master functions in the two schemes read
\begin{align}
\bS^{(i,j)}=
\begin{cases}
\bs^{\st}\bK^{j}(\bI+\bM)^{-1}\bK^{i}\br , \quad \mbox{AKNS case},\\
\bs^{\st}\bL^{j}\bM^{-1}\bK^{i}\br , \quad \mbox{KP case}.
\end{cases}
\end{align}
\end{remark}
Detailed proofs of the above relations and scheme classifications can be found in \cite{HSZ2025,ZMZ2026}. In the following derivation, we proceed from the general DES \eqref{DES} without specifying the scheme, and return to the two schemes in Section 3 when constructing explicit solutions.

To derive the symmetric discrete Mikhailov model, we define matrix-valued variables from the master functions 
\begin{align}
	\label{uvw-def}
	&\bu=\bS^{(0,0)}=\begin{pmatrix}
		u_{1}&u_{2}\\
		u_{3}&u_{4}
	\end{pmatrix}, \quad \bv=\bI-\bS^{(-1,0)}=\begin{pmatrix}
		v_{1}&v_{2}\\
		v_{3}&v_{4}
	\end{pmatrix},\quad \bw=\bI+\bS^{(0,-1)},
\end{align}
where $u_i, v_{i}~(i=1,2,3,4)$ are scalar functions. Directly from relations \eqref{Sij-re=1} and \eqref{sh-Sij}, one can obtain the following identities at the matrix level.
\begin{prop}
The matrix functions $\bu$, $\bv$ and $\bw$ introduced in \eqref{uvw-def} satisfy
\begin{subequations}
\label{re-uvw}
\begin{align}
\label{vw}
&\bv\bw=\bI,\\
\label{h-shift-uvw}
&\bv\ba\wh{\bw}=\wh{\bv}\ba\bw=\ba-q(\wh{\bu}-\bu),\\
\label{t-shift-uvw3}
&p\bv\wt{\bw}=p\bI+\bu\ba-\ba\wt{\bu},\\
\label{t-shift-uvw4}
&p\wt{\bv}\bw=p\bI+\ba\bu-\wt{\bu}\ba,
\end{align}
\end{subequations}
which further imply
\begin{subequations}
\label{re-u}
\begin{align}
\label{h-shift-u}
&(\ba-q(\wh{\bu}-\bu))^2 =\bI,\\
\label{t-shift-u}
&(p\bI+\bu\ba-\ba\wt{\bu})(p\bI +\ba\bu-\wt{\bu}\ba)=p^2\bI,\\
\label{u-closed}
&(\ba-q(\wh{\wt{\bu}}-\wt{\bu}))(p\bI+\ba\wh{\bu}-\wh{\wt{\bu}}\ba)
=(p\bI+\ba\bu-\wt{\bu}\ba)(\ba-q(\wh{\bu}-\bu)).
\end{align}
\end{subequations}
\end{prop}
The proof of these matrix identities follows the same procedure as in Ref. \cite{ZMZ2026}. To derive scalar-form symmetric discrete Mikhailov models, we expand the matrix relations into scalar form and introduce the dependent variable $\nu:=\mathrm{i}v_{2}/v_{4}$. Under the normalization $|\bv|=1$ (justified in \cite{ZMZ2026}), $\bw$ admits the explicit form
\begin{align}
\label{vw-def1}
\bw=\bI+\bS^{(0,-1)}=\begin{pmatrix}
v_{4}&-v_{2}\\
-v_{3}&v_{1}
\end{pmatrix}.
\end{align}
We then arrive at the following scalar relations, which serve as the starting point for constructing symmetric discrete Mikhailov models.
\begin{prop}
Define auxiliary quantities
\begin{align}
&A_{1}:=1-q(\wh{u}_{1}-u_{1}),\quad A_{2}:=q(\wh{u}_{2}-u_{2}),\quad A_{3}:=-q(\wh{u}_{3}-u_{3}),\\
&B_{1}:=\wt{u}_{1}-u_{1},\quad B_{2}:=\wt{u}_{2}+u_{2}, \quad B_{3}:=-\wt{u}_{3}-u_{3}.
\end{align}
The following identities hold
\begin{subequations}
\begin{align}
\label{Ai-relation}
&A_{1}^2-A_{2}A_{3}=1,\\
\label{A2}
&A_{2}=A_{3}\nu\wh{\nu}-\mathrm{i}A_{1}(\wh{\nu}+\nu),\\
\label{Bi-relation}
&(p-B_{1})^{2}=p^{2}+B_{2}B_{3},\\
\label{t-Biv5-1}
&B_{2}=\mathrm{i}(p-B_{1})(\nu-\wt{\nu})-B_{3}\nu\wt{\nu},
\end{align}
\end{subequations}
from which it follows that
\begin{subequations}
\label{A1B1}
\begin{align}
\label{A1}
&2A_{1}=\mathrm{i}q(\wh{u}_{3}-u_{3})(\wh{\nu}+\nu)+\big(4-q^2(u_{3}-\wh{u}_{3})^2(\nu-\wh{\nu})^2\big)^{\frac{1}{2}},\\
\label{B1}
&2(p-B_{1})=\mathrm{i}(u_{3}+\wt{u}_{3})(\wt{\nu}-\nu)+\big(4p^2-(u_{3}+\wt{u}_{3})^2(\nu+\wt{\nu})^2 \big)^{\frac{1}{2}}.
\end{align}
\end{subequations}
\end{prop}

\subsection{Symmetric discrete Mikhailov models}

The discrete Mikhailov model derived in Ref. \cite{ZMZ2026} lacks conjugate symmetry, which prevents direct reduction to the dFL equation. 
In this subsection, we resolve this issue by exploiting the equivalent representations of $u_{1}$-related terms in the compatibility condition, 
leading to four families of fully symmetric discrete Mikhailov models.

We begin with the $(2,1)$-component of the matrix equation \eqref{u-closed}, i.e.,
\begin{align}
	\label{u3u1}
	(\wh{u}_{3}+\wt{u}_{3})(1/q+p+u_{1}-\wh{\wt{u}}_{1})+(u_{3}+\wh{\wt{u}}_{3})(1/q-p+\wt{u}_{1}-\wh{u}_{1})=0.
\end{align}
Combining this with the Miura-type relations \eqref{A2} and \eqref{t-Biv5-1}, and substituting the identity
\begin{align}
	A_{2}/q+B_{2}=\wh{B}_{2}-\wt{A}_{2}/q,
\end{align}
we obtain
\begin{align}
	\label{vu1}
	(\wh{\nu}+\wt{\nu})(1/q+p+u_{1}-\wh{\wt{u}}_{1}+\mathrm{i}u_{3}\nu-\mathrm{i}\wh{\wt{u}}_{3}\wh{\wt{\nu}})+(\nu+\wh{\wt{\nu}})
(1/q-p+\wt{u}_{1}-\wh{u}_{1}+\mathrm{i}\wt{u}_{3}\wt{\nu}-\mathrm{i}\wh{u}_{3}\wh{\nu})=0.
\end{align}

Equations \eqref{u3u1} and \eqref{vu1} both feature $u_{1}$-dependent terms. To eliminate $u_{1}$ and close the system in $u_{3}$ and $\nu$, 
we note that each of the combinations $1/q+p+u_{1}-\wh{\wt{u}}_{1}$ and $1/q-p+\wt{u}_{1}-\wh{u}_{1}$ admits two equivalent representations in terms of 
$A_{1}$ and $(p-B_{1})$:
\begin{align}
1/q+p+u_{1}-\wh{\wt{u}}_{1}=\begin{cases}
		A_{1}/q+(p-\wh{B}_{1}),\\
		\wt{A}_{1}/q+(p-B_{1}),
	\end{cases}
1/q-p+\wt{u}_{1}-\wh{u}_{1}=\begin{cases}
	A_{1}/q-(p-B_{1}),\\
	\wt{A}_{1}/q-(p-\wh{B}_{1}).
\end{cases}
\end{align}
Pairwise combinations of these equivalent relations yield four distinct families of symmetric discrete Mikhailov models for $(\mu,\nu):=(u_{3},\mathrm{i}v_{2}/v_{4})$, as summarized below,
where $A_{1}$ and $B_{1}$ should be replaced by the expressions given in \eqref{A1B1}, $P$ and $Q$ are defined as
\begin{align}
\label{PQ}
P:=\big(4p^2-(\mu+\wt{\mu})^2(\nu+\wt{\nu})^2\big)^{\frac{1}{2}},\quad Q:=\big(4/q^2-(\mu-\wh{\mu})^2(\nu-\wh{\nu})^2\big)^{\frac{1}{2}}.
\end{align}

\noindent$\bullet$~~\textbf{Family 1}: Taking $1/q+p+u_{1}-\wh{\wt{u}}_{1}=A_{1}/q+(p-\wh{B}_{1})$ and $1/q-p+\wt{u}_{1}-\wh{u}_{1}=A_{1}/q-(p-B_{1})$ yields
\begin{subequations}
	\label{dFL1}
	\begin{align}
		\label{dFL1-1}
		&(\wh{\mu}+\wt{\mu})\big((\nu+\wh{\wt{\nu}})(\wh{\mu}+\wh{\wt{\mu}})-\mathrm{i}\wh{P}-\mathrm{i}Q\big)
		-(\mu+\wh{\wt{\mu}})\big((\wh{\nu}+\wt{\nu})(\mu+\wt{\mu})+\mathrm{i}Q-\mathrm{i}P\big)=0,\\
		\label{dFL1-2}
		&(\wh{\nu}+\wt{\nu})\big((\mu+\wh{\wt{\mu}})(\wh{\nu}+\wh{\wt{\nu}})+\mathrm{i}\wh{P}+\mathrm{i}Q\big)
		-(\nu+\wh{\wt{\nu}})\big((\wh{\mu}+\wt{\mu})(\nu+\wt{\nu})-\mathrm{i}Q+\mathrm{i}P\big)=0.
	\end{align}
\end{subequations}	
\noindent$\bullet$~~\textbf{Family 2}: Taking $1/q+p+u_{1}-\wh{\wt{u}}_{1}=A_{1}/q+(p-\wh{B}_{1})$ and $1/q-p+\wt{u}_{1}-\wh{u}_{1}=\wt{A}_{1}/q-(p-\wh{B}_{1})$ yields
\begin{subequations}
	\label{dFL2}
	\begin{align}
		\label{dFL2-1}
		&(\wh{\mu}+\wt{\mu})\big((\nu+\wh{\wt{\nu}})(\wh{\mu}-\mu)-\mathrm{i}\wh{P}-\mathrm{i}Q\big)
		+(\mu+\wh{\wt{\mu}})\big((\wh{\nu}+\wt{\nu})(\wh{\wt{\mu}}-\wt{\mu})-\mathrm{i}\wt{Q}+\mathrm{i}\wh{P}\big)=0,\\
		\label{dFL2-2}
		&(\wh{\nu}+\wt{\nu})\big((\mu+\wh{\wt{\mu}})(\wh{\nu}-\nu)+\mathrm{i}\wh{P}+\mathrm{i}Q\big)
		+(\nu+\wh{\wt{\nu}})\big((\wh{\mu}+\wt{\mu})(\wh{\wt{\nu}}-\wt{\nu})+\mathrm{i}\wt{Q}-\mathrm{i}\wh{P}\big)=0.
	\end{align}
\end{subequations}	
\noindent$\bullet$~~\textbf{Family 3}: Taking $1/q+p+u_{1}-\wh{\wt{u}}_{1}=\wt{A}_{1}/q+(p-B_{1})$ and $1/q-p+\wt{u}_{1}-\wh{u}_{1}=A_{1}/q-(p-B_{1})$ yields
	\begin{subequations}
		\label{dFL3}
	\begin{align}
		\label{dFL3-1}
		&(\wh{\mu}+\wt{\mu})\big((\nu+\wh{\wt{\nu}})(\wh{\wt{\mu}}-\wt{\mu})-\mathrm{i}P-\mathrm{i}\wt{Q}\big)
		+(\mu+\wh{\wt{\mu}})\big((\wh{\nu}+\wt{\nu})(\wh{\mu}-\mu)-\mathrm{i}Q+\mathrm{i}P\big)=0,\\
		\label{dFL3-2}
		&(\wh{\nu}+\wt{\nu})\big((\mu+\wh{\wt{\mu}})(\wh{\wt{\nu}}-\wt{\nu})+\mathrm{i}P+\mathrm{i}\wt{Q}\big)
		+(\nu+\wh{\wt{\nu}})\big((\wh{\mu}+\wt{\mu})(\wh{\nu}-\nu)+\mathrm{i}Q-\mathrm{i}P\big)=0.
	\end{align}
\end{subequations}	
\noindent$\bullet$~~\textbf{Family 4}: Taking $1/q+p+u_{1}-\wh{\wt{u}}_{1}=\wt{A}_{1}/q+(p-B_{1})$ and $1/q-p+\wt{u}_{1}-\wh{u}_{1}=\wt{A}_{1}/q-(p-\wh{B}_{1})$ yields
\begin{subequations}
	\label{dFL4}
	\begin{align}
		\label{dFL4-1}
		&(\wh{\mu}+\wt{\mu})\big((\nu+\wh{\wt{\nu}})(\mu+\wt{\mu})+\mathrm{i}P+\mathrm{i}\wt{Q}\big)
		-(\mu+\wh{\wt{\mu}})\big((\wh{\nu}+\wt{\nu})(\wh{\mu}+\wh{\wt{\mu}})-\mathrm{i}\wt{Q}+\mathrm{i}\wh{P}\big)=0,\\
		\label{dFL4-2}
		&(\wh{\nu}+\wt{\nu})\big((\mu+\wh{\wt{\mu}})(\nu+\wt{\nu})-\mathrm{i}P-\mathrm{i}\wt{Q}\big)
		-(\nu+\wh{\wt{\nu}})\big((\wh{\mu}+\wt{\mu})(\wh{\nu}+\wh{\wt{\nu}})+\mathrm{i}\wt{Q}-\mathrm{i}\wh{P}\big)=0.
	\end{align}
\end{subequations}
The four families constructed above are all symmetric discrete Mikhailov models that can be regarded as quad systems involving square-root edge variables, each of which admits 
a multi-valued evolution from prescribed initial data
$\{(\mu,\nu), (\wt{\mu},\wt{\nu}), (\wh{\mu},\wh{\nu})\}$ (see \cite{AN2014}). Owing to their conjugate symmetry, they admit local reductions to dFL
equations. Appropriate pairwise recombination of the lattice equations further yields combined systems that support nonlocal reductions,
corresponding to nonlocal dFL equations. The solution analysis for both local and nonlocal dFL equations is presented in Section 5.
It is worth noting that the connection (see \cite{MWZ2026}, Theorem 2.1) between the discrete four-component MTM \cite{MWZ2026} and the 
discrete Mikhailov model established in \cite{ZMZ2026} remains valid for the models derived in this paper.

\begin{remark}
In addition to the dependent variable identification $(\mu,\nu):=(u_{3},\mathrm{i}v_{2}/v_{4})$ used in the main derivation, the alternative choice
\begin{align}
(\mu,\nu):=(\mathrm{i}v_{3}/v_{1},u_{2})
\end{align}
also yields valid solutions to all four families of symmetric discrete Mikhailov models \eqref{dFL1}-\eqref{dFL4}.
\end{remark}

\begin{remark}
The asymmetric discrete Mikhailov model proposed in Ref. \cite{ZMZ2026} does not belong to any of the four symmetric families derived above. 
It is essentially a formulation that cross-pairs equations from two different supplementary
equivalent forms: one equation is taken from the first supplementary form and the other from the second supplementary form. 
This mixed matching across different forms is the root cause of its absence of conjugate symmetry. By contrast, the four symmetric formulations constructed herein remove this structural asymmetry and support symmetry reductions. The complete expressions and derivations of the two supplementary forms are presented in Appendix A.
\end{remark}

\begin{remark}
The procedure used to construct the symmetric discrete Mikhailov models can be extended to the subsystem for the field pair $(u_{2},u_{3})$, and yields four families of discrete negative-order AKNS models as a natural by-product. Ref. \cite{Z2016} derived a single formulation of the discrete negative-order AKNS equation via the Cauchy matrix approach. In comparison, we obtain here four distinct families of discrete negative-order AKNS models, which are more compact in form and all exhibit inherent conjugate symmetry.
	
For convenience, we define the auxiliary quantities
\begin{align}
\Psi_{1}=\big(p^2-(\wt{u}_{2}+u_{2})(\wt{u}_{3}+u_{3})\big)^{\frac{1}{2}},\quad\Psi_{2}=\big(1/q^2-(\wh{u}_{2}-u_{2})(\wh{u}_{3}-u_{3})\big)^{\frac{1}{2}}.
\end{align}
The four families of discrete negative-order AKNS models are listed as follows.
\begin{enumerate}
\item
\begin{subequations}
\begin{align}
&(\wh{u}_{3}+\wt{u}_{3})(\Psi_{2}+\wh{\Psi}_{1})+(u_{3}+\wh{\wt{u}}_{3})(\Psi_{2}-\Psi_{1})=0,\\
&(\wh{u}_{2}+\wt{u}_{2})(\Psi_{2}+\wh{\Psi}_{1})+(u_{2}+\wh{\wt{u}}_{2})(\Psi_{2}-\Psi_{1})=0.
\end{align}
\end{subequations}
\item
\begin{subequations}
\begin{align}
&(\wh{u}_{3}+\wt{u}_{3})(\Psi_{2}+\wh{\Psi}_{1})+(u_{3}+\wh{\wt{u}}_{3})(\wt{\Psi}_{2}-\wh{\Psi}_{1})=0,\\
&(\wh{u}_{2}+\wt{u}_{2})(\Psi_{2}+\wh{\Psi}_{1})+(u_{2}+\wh{\wt{u}}_{2})(\wt{\Psi}_{2}-\wh{\Psi}_{1})=0.
\end{align}
\end{subequations}
\item
\begin{subequations}
\begin{align}
&(\wh{u}_{3}+\wt{u}_{3})(\wt{\Psi}_{2}+\Psi_{1})+(u_{3}+\wh{\wt{u}}_{3})(\Psi_{2}-\Psi_{1})=0,\\
&(\wh{u}_{2}+\wt{u}_{2})(\wt{\Psi}_{2}+\Psi_{1})+(u_{2}+\wh{\wt{u}}_{2})(\Psi_{2}-\Psi_{1})=0.
\end{align}
\end{subequations}
\item
\begin{subequations}
\begin{align}
&(\wh{u}_{3}+\wt{u}_{3})(\wt{\Psi}_{2}+\Psi_{1})+(u_{3}+\wh{\wt{u}}_{3})(\wt{\Psi}_{2}-\wh{\Psi}_{1})=0,\\
&(\wh{u}_{2}+\wt{u}_{2})(\wt{\Psi}_{2}+\Psi_{1})+(u_{2}+\wh{\wt{u}}_{2})(\wt{\Psi}_{2}-\wh{\Psi}_{1})=0.
\end{align}
\end{subequations}
\end{enumerate}
\end{remark}

\section{Explicit solutions}\label{sec-3}

The symmetric discrete Mikhailov models admit the same underlying Cauchy matrix structure as the original model in Ref. \cite{ZMZ2026}. 
Accordingly, the AKNS-type and KP-type solution frameworks are still applicable, with the construction logic for plane-wave factors, 
solutions to the Sylvester equation and master function expressions unchanged. In this section, we briefly present the core forms of 
these two solution schemes, which underpin the dFL solutions derived in Section 5. Full derivations for the associated matrix 
structures are provided in \cite{ZMZ2026}.

We first define the discrete plane wave factors:
\begin{subequations}
\label{plane-def}
\begin{align}
\label{plane-rho-def}
&\rho_{\iota}(z)=\left( 1+(-1)^{\iota-1}\frac{z}{p}\right)^{n}
\left(1+ (-1)^{\iota-1}\frac{1}{qz}\right)^{m}\rho^{0}_{\iota}(z),\\
&\sigma_{\iota}(z)=\left( 1+(-1)^{\iota-1}\frac{z}{p}\right)^{-n}
\left(1+ (-1)^{\iota-1}\frac{1}{qz}\right)^{-m}\sigma^{0}_{\iota}(z),
\end{align}
\end{subequations}
where $\rho^{0}_{\iota}(z),~\sigma^{0}_{\iota}(z),~(\iota=1,2)$ are constant phase parameters independent of $n$ and $m$.
	
\subsection{The AKNS-type scheme}\label{sec-3-1}
For the AKNS-type scheme ($\bL=\bK, N=2 \mathcal{N}$ ), we introduce block-structured matrices
\begin{align}
	\label{Sym-KM12rts-def}
	\bK=\begin{pmatrix}
		\bK_{1}&\\
		&\bK_{2}
	\end{pmatrix}, \quad \bM=\begin{pmatrix}
		&\bM_{1}\\
		\bM_{2}&
	\end{pmatrix}, \quad \br=\begin{pmatrix}
		\br_{1}&\\
		&\br_{2}
	\end{pmatrix}, \quad \bs=\begin{pmatrix}
		& \bs_{1} \\
		\bs_{2} &
	\end{pmatrix},
\end{align}	
where $\bK_{\iota}, \bM_{\iota} \in \mathbb{C}^{\mathcal{N}\times \mathcal{N}}$, and $\br_{\iota},\bs_{\iota}$ are 
$\mathcal{N}$-dimensional column vectors for $\iota=1,2$. The DES \eqref{DES} decouples into
\begin{subequations}
\label{DES-sym-ent}
\begin{align}
\label{sym-SE-12}
& \bK_{1}\bM_{1}-\bM_{1}\bK_{2}=\br_{1}\bs^{\st}_{2},\quad \bK_{2}\bM_{2}-\bM_{2}\bK_{1}=\br_{2}\bs^{\st}_{1}, \\
\label{sym-shift-r1}
& p\wt{\br}_{\iota}=p\br_{\iota}+(-1)^{\iota-1}\bK_{\iota}\br_{\iota},\quad q\wh{\br}_{\iota}=q\br_{\iota}+(-1)^{\iota-1}\bK^{-1}_{\iota}\br_{\iota}, \\
\label{sym-shift-s1}
& p\wt{\bs}_{\iota}+(-1)^{\iota}\bK^{\st}_{\iota}~\wt{\bs}_{\iota}=p\bs_{\iota},\quad q\wh{\bs}_{\iota}+(-1)^{\iota}(\bK_{\iota}^{-1})^{\st}~\wh{\bs}_{\iota}=q\bs_{\iota},
\end{align}
\end{subequations}
under the assumption that $\bK_1$ and $\bK_2$ have no common eigenvalues.

The dependent variables $(\mu,\nu)$ of the symmetric discrete Mikhailov models are expressed as
\begin{align}
\label{sym-FL-solu}
(\mu,\nu)=\left( \bs^{\st}_{1}(\bI-\bM_{1}\bM_{2})^{-1}\br_{1},
\frac{-\mathrm{i}\bs^{\st}_{2}(\bI-\bM_{2}\bM_{1})^{-1}\bK_{2}^{-1}\br_{2}}{1+\bs^{\st}_{1}
\bM_{1}(\bI-\bM_{2}\bM_{1})^{-1}\bK_{2}^{-1}\br_{2}}\right).
\end{align}
	
\vskip 5pt
\noindent
\textbf{Soliton solutions}. For diagonal $\bK_{1}=\text{diag}(k_{1},k_{2},\cdots,k_{\mathcal{N}})$ and $\bK_{2}=\text{diag}(l_{1},l_{2},\cdots,l_{\mathcal{N}})$ with $k_i\neq l_j$, the soliton solution is constructed with
\begin{subequations}
\label{sym-SE-solu}
\begin{align}
&\br_{1}=(\rho_{1}(k_{1}),\rho_{1}(k_{2}),\cdots,\rho_{1}(k_{\mathcal{N}}))^{\st}, \quad \br_{2}=(\rho_{2}(l_{1}),\rho_{2}(l_{2}),\cdots,\rho_{2}(l_{\mathcal{N}}))^{\st},\\
&\bs_{1}=(\sigma_{2}(k_{1}),\sigma_{2}(k_{2}),\cdots,\sigma_{2}(k_{\mathcal{N}}))^{\st}, \quad \bs_{2}=(\sigma_{1}(l_{1}),\sigma_{1}(l_{2}),\cdots,\sigma_{1}(l_{\mathcal{N}}))^{\st},\\
\label{AKNS-M1-def}
&\bM_{1}=(M_{1,ij})_{\mathcal{N}\times \mathcal{N}},\quad M_{1,ij}=\frac{\rho_{1}(k_{i})\sigma_{1}(l_{j})}{k_{i}-l_{j}},\\
\label{AKNS-M2-def}
&\bM_{2}=(M_{2,ij})_{\mathcal{N}\times \mathcal{N}},\quad M_{2,ij}=\frac{\rho_{2}(l_{i})\sigma_{2}(k_{j})}{l_{i}-k_{j}}.
\end{align}
\end{subequations}
	
\vskip 5pt
\noindent
\textbf{Multiple-pole solutions}. Multiple-pole solutions arise from Jordan-block structures of $\bK_1$ and $\bK_2$. We consider two canonical cases:
	
\noindent$\bullet$\hspace{1em}\textit{Case-I}. $\bK_{1}=\bJ[k,1]$~(standard Jordan block) and $\bK_{2}=\bJ[l,-1]$~(Jordan-block-like matrix with off-diagonal element $-1$), where
\begin{align}
\label{J-def}
&\bJ[z,\delta]:=\begin{pmatrix}
z&0&0&\cdots&0\\
\delta&z&0&\cdots&0\\
0&\delta&z&\cdots&0\\
\vdots&\vdots&\vdots&\ddots&\vdots\\
0&0&0&\cdots&z
\end{pmatrix}_{\mathcal{N}\times \mathcal{N}}.
\end{align}
In this situation, we have
\begin{subequations}
\label{JS-sym}
\begin{align}
&\br_{1}=\bF_{k}[\rho_{1}(k)]\bar{\be}_{\mathcal{N}},\quad \br_{2}=\bF_{-l}[\rho_{2}(l)]\bar{\be}_{\mathcal{N}}, \\
&\bs_{1}=\bH_{k}[\sigma_{2}(k)]\bar{\be}_{\mathcal{N}},\quad
\bs_{2}=\bH_{-l}[\sigma_{1}(l)]\bar{\be}_{\mathcal{N}},\\
&\bM_{1}=\bF_{k}[\rho_{1}(k)]\bar{\bG}\bH_{-l}[\sigma_{1}(l)],\quad
\bM_{2}=-\bF_{-l}[\rho_{2}(l)]\bar{\bG}\bH_{k}[\sigma_{2}(k)], \\
&\bar{\bG}=(\bar{G}_{i,j})_{\mathcal{N}\times \mathcal{N}},\quad \bar{G}_{i,j}=\frac{(-1)^{i+j}
\mathrm{C}^{i-1}_{i+j-2}}{(k-l)^{i+j-1}},\quad \bar{\be}_{\mathcal{N}}=(\underbrace{1, 0, \cdots, 0}_{\mathcal{N}\text{-dimensional}})^{\st},
\end{align}
\end{subequations}
where $\mathrm{C}^{i}_{j}=\frac{j!}{i!(j-i)!}$ for $j\geq i$.
	
\noindent$\bullet$\hspace{1em}\textit{Case-II}. $\bK_{1}=\bJ[k,1]$ and $\bK^{\st}_{2}=\bJ[l,1]$. In this case, we have
\begin{subequations}
\begin{align}
&\br_{1}=\bF_{k}[\rho_{1}(k)]\bar{\be}_{\mathcal{N}},\quad \br_{2}=\bH_{l}[\rho_{2}(l)]\bar{\be}_{\mathcal{N}},\\
&\bs_{1}=\bH_{k}[\sigma_{2}(k)]\bar{\be}_{\mathcal{N}},\quad
\bs_{2}=\bF_{l}[\sigma_{1}(l)]\bar{\be}_{\mathcal{N}},\\
&\bM_{1}=\bF_{k}[\rho_{1}(k)]\bG_{1}\bF^{\st}_{l}[\sigma_{1}(l)],
\quad\bM_{2}=\bH_{l}[\rho_{2}(l)]\bG_{2}\bH_{k}[\sigma_{2}(k)],\\
&\bG_{1}=(g^{(1)}_{i,j})_{\mathcal{N}\times \mathcal{N}},\quad g^{(1)}_{i,j}=\frac{(-1)^{i-1}\mathrm{C}^{i-1}_{i+j-2}}{(k-l)^{i+j-1}},\\
&\bG_{2}=(g^{(2)}_{i,j})_{\mathcal{N}\times \mathcal{N}},\quad g^{(2)}_{i,j}=\frac{(-1)^{i-1}\mathrm{C}^{i-1}_{i+j-2}}{(l-k)^{i+j-1}}.
\end{align}
\end{subequations}
In both cases, the vector and matrix functions are constructed via the lower triangular Toeplitz matrix $\bF_{\delta z}[f(z)]$
and skew triangular Hankel matrix $\bH_{\delta z}[f(z)]$, whose definitions are
\begin{subequations}
\label{FH-def}
\begin{align}
\label{F-def}
&\bF_{\delta z}[f(z)]=\begin{pmatrix}
f(z)&0&0&\cdots&0\\
\partial_{\delta z}f(z)&f(z)&0&\cdots&0\\
\frac{\partial^{2}_{\delta z}f(z)}{2!}&\partial_{\delta z}f(z)&f(z)
&\cdots&0\\
\vdots&\vdots&\vdots&\ddots&\vdots\\
\frac{\partial^{\mathcal{N}-1}_{\delta z}f(z)}{(\mathcal{N}-1)!}
&\frac{\partial^{\mathcal{N}-2}_{\delta z}f(z)}{(\mathcal{N}-2)!}
&\frac{\partial^{\mathcal{N}-3}_{\delta z}f(z)}{(\mathcal{N}-3)!}&\cdots&f(z)
\end{pmatrix}_{\mathcal{N}\times \mathcal{N}},\\
\label{H-def}
&\bH_{\delta z}[f(z)]=
\begin{pmatrix}
\frac{\partial^{\mathcal{N}-1}_{\delta z}f(z)}{(\mathcal{N}-1)!} & \cdots 
& \frac{\partial^{2}_{\delta z}f(z)}{2!} & \partial_{\delta z}f(z) & f(z)\\
\frac{\partial^{\mathcal{N}-2}_{\delta z}f(z)}{(\mathcal{N}-2)!} 
& \cdots & \partial_{\delta z}f(z) & f(z) & 0\\
\frac{\partial^{\mathcal{N}-3}_{\delta z}f(z)}{(\mathcal{N}-3)!} & \cdots & f(z) & 0 & 0\\
\vdots & \vdots & \vdots & \ddots &\ddots\\
f(z) & \cdots & 0 & 0 & 0
\end{pmatrix}_{\mathcal{\mathcal{N}}\times \mathcal{\mathcal{N}}}.
\end{align}
\end{subequations}
Substituting the Jordan-block forms into \eqref{sym-FL-solu} yields the $\mathcal{N}$-th order multiple-pole solution.

\subsection{The KP-type scheme}\label{sec-3-2}

For the KP-type scheme~($\bL\neq \bK,\bC=\bm 0$), we decompose $\br=(\br_{1},\br_{2})$, $\bs=(\bs_{1},\bs_{2})$ and $\bM=\bM_{1}+\bM_{2}$, 
leading to two decoupled scalar KP-type Cauchy matrix systems
\begin{subequations}
\label{asym-shift1}
\begin{align}
\label{asym-SE12}
& \bK\bM_{\iota}-\bM_{\iota}\bL=\br_{\iota}\bs^{\st}_{\iota}, \\
\label{asym-shift-r1}
& p\wt{\br}_{\iota}=p\br_{\iota}+(-1)^{\iota-1}\bK\br_{\iota},\quad q\wh{\br}_{\iota}=q\br_{\iota}+(-1)^{\iota-1}\bK^{-1}\br_{\iota}, \\
\label{asym-shift-s1}
& p\wt{\bs}_{\iota}+(-1)^{\iota-1}\bL^{\st}~\wt{\bs}_{\iota}=p\bs_{\iota},\quad
q\wh{\bs}_{\iota}+(-1)^{\iota-1}(\bL^{-1})^{\st}~\wh{\bs}_{\iota}=q\bs_{\iota},
\end{align}
\end{subequations}
and the dependent variables are given by
\begin{align}
\label{asym-FL-solu}
\left( \mu,\nu\right)= \left(\bs^{\st}_{2}(\bM_{1}+\bM_{2})^{-1}\br_{1},
\frac{-\mathrm{i}\bs^{\st}_{1}
(\bM_{1}+\bM_{2})^{-1}\bK^{-1}\br_{2}}{1-\bs^{\st}_{2}(\bM_{1}
+\bM_{2})^{-1}\bK^{-1}\br_{2}}\right).
\end{align}

\vskip 5pt
\noindent
\textbf{Soliton solutions}. Functions \eqref{asym-FL-solu} provide soliton solutions to the symmetric discrete 
Mikhailov models \eqref{dFL1}-\eqref{dFL4} through the KP-type Cauchy matrix scheme, in which
\begin{subequations}
\label{asym-SE-solu}
\begin{align}
& \bK=\text{diag}(k_{1},k_{2},\cdots,k_{N}),\quad\bL=\text{diag}(l_{1},l_{2},\cdots,l_{N}),\\
\label{asym-M1M2-decom}
&\br_{\iota}=(\rho_{\iota}(k_{1}),\rho_{\iota}(k_{2}),\cdots,\rho_{\iota}(k_{N}))^{\st},\quad \bs_{\iota}=(\sigma_{\iota}(l_{1}),\sigma_{\iota}(l_{2}),\cdots,\sigma_{\iota}(l_{N}))^{\st},\\
&\bM_{\iota}=(M_{\iota,ij})_{N\times N},\quad M_{\iota,ij}=\frac{\rho_{\iota}(k_{i})\sigma_{\iota}(l_{j})}{k_{i}-l_{j}},
\end{align}
\end{subequations}
with $k_i\neq l_j,~(i,j=1,2,\ldots,N$), where $\rho_{\iota},~\sigma_{\iota}~(\iota=1,2)$ are defined in \eqref{plane-def}.
	
\vskip 5pt
\noindent
\textbf{Multiple-pole solutions}.
\begin{subequations}
\begin{align}
&\bK=\bJ[k,1],\quad\bL^{\st}=\bJ[l,1],\\
&\br_{\iota}=\bF_{k}[\rho_{\iota}(k)]\bar{\be}_{N},\quad
\bs_{\iota}=\bF_{l}[\sigma_{\iota}(l)]\bar{\be}_{N}, \quad \bar{\be}_{N}
=(\underbrace{1, 0, \cdots, 0}_{N\text{-dimensional}})^{\st}, \\
&\bM=\bM_{1}+\bM_{2},\quad\bM_{\iota}
=\bF_{k}[\rho_{\iota}(k)]\bar{\bG}\bF^{\st}_{l}[\sigma_{\iota}(l)],
\quad \iota=1,2,\\
&\bar{\bG}=(\bar{G}_{i,j})_{N\times N},\quad \bar{G}_{i,j}=\frac{(-1)^{i-1}\mathrm{C}^{i-1}_{i+j-2}}{(k-l)^{i+j-1}}.
\end{align}
\end{subequations}

\section{Continuum limits}\label{sec-4}
In this section, we verify that all four families of symmetric discrete Mikhailov models recover the continuous Mikhailov model via 
two-step continuum limits. The limit procedure follows the same scaling as in Ref. \cite{ZMZ2026}; 
we present only the resulting semi-discrete and continuous systems, along with their corresponding Cauchy matrix structures, 
and omit the detailed Taylor expansion derivations. For the solutions to the semi-discrete and continuous Mikhailov model, 
one can refer to \cite{ZMZ2026}.

\subsection{Semi-discrete Mikhailov model}\label{sec-4-1}
In the first step, we take $q\rightarrow\infty$, $m\rightarrow\infty$
with the continuous spatial coordinate $x=m/q$ kept finite. The $m$-direction shift is expanded as
\begin{align}
\wh{f}=f(n,x+1/q)=f+f_{x}/q+f_{xx}/(2q^2)+\cdots.
\end{align}
Inserting them into \eqref{A1}, \eqref{u3u1} and \eqref{vu1}, respectively, then from the leading term (in terms of $1/q$) we obtain a semi-discrete model
\begin{subequations}
\label{sd-u1u3v}
\begin{align}
&u_{1,x}+\mathrm{i}\nu u_{3,x}=0,\\
&(u_{3}+\wt{u}_{3})(u_{1,x}+\wt{u}_{1,x}-2)+(\wt{u}_{3,x}-u_{3,x})(p+u_{1}-\wt{u}_{1})=0,\\
&(\nu+\wt{\nu})(u_{1,x}+\wt{u}_{1,x}-2+\mathrm{i}u_{3,x}\nu+\mathrm{i}\wt{u}_{3,x}\wt{\nu})+(\wt{\nu}_{x}-\nu_{x})(p+u_{1}-\wt{u}_{1})
+\mathrm{i}(u_{3}+\wt{u}_{3})(\nu\wt{\nu}_{x}+\wt{\nu}\nu_{x})=0.
\end{align}
\end{subequations}
It is observed that this semi-discrete model also contains a term involving $u_{1}$, which we need to eliminate. 
With the aid of equations \eqref{B1}, the semi-discrete model \eqref{sd-u1u3v} simplifies to a form that involves only $\mu$ and $\nu$ with $\mu:=u_{3}$:
\begin{subequations}
\label{sd-u3v}
\begin{align}
&(\mu+\wt{\mu})\big(\mathrm{i}(\mu_{x}+\wt{\mu}_{x})(\nu+\wt{\nu})+4\big)+(\mu_{x}-\wt{\mu}_{x})P^{'}=0,\\ 
&(\nu+\wt{\nu})\big(\mathrm{i}(\nu_{x}+\wt{\nu}_{x})(\mu+\wt{\mu})-4\big)-(\nu_{x}-\wt{\nu}_{x})P^{'}=0, 			
\end{align}
where
\begin{align}
P^{'}:=P^{'}(n,x)=\big(4p^2-(\mu+\wt{\mu})^2(\nu+\wt{\nu})^2\big)^{\frac{1}{2}}.
\end{align}
\end{subequations}
We refer to \eqref{sd-u3v} as the semi-discrete Mikhailov model.

Under this limit, the discrete plane wave factors \eqref{plane-def} become
\begin{subequations}
\label{semi-pwf}
\begin{align}
&\rho_{\iota}(z)=\left(1+ (-1)^{\iota-1}\frac{z}{p}\right)^{n}
e^{(-1)^{\iota-1}z^{-1}x}\rho^{0}_{\iota}(z), \\
&\sigma_{\iota}(z)=\left(1+ (-1)^{\iota-1}\frac{z}{p}\right)^{-n}
e^{(-1)^{\iota}z^{-1}x}\sigma^{0}_{\iota}(z).
\end{align}
\end{subequations}
	

\subsection{Continuous Mikhailov model}
In the second step, we take $p\rightarrow\infty$, $n\rightarrow\infty$, with the continuous temporal coordinate $t=n/p$
kept finite. The $n$-direction shift is expanded as
\begin{align}
\wt{f}=f(t+1/p,x)=f+f_{t}/p+f_{tt}/(2p^2)+\cdots.
\end{align}
Substituting into the semi-discrete Mikhailov model \eqref{sd-u3v} and retaining the leading-order terms, we recover the continuous Mikhailov model:
\begin{subequations}
\label{FL}
\begin{align}
\label{FL-1}
&\mu_{xt}-4\mu-4\mathrm{i}\mu\nu \mu_{x}=0,\\
\label{FL-2}
&\nu_{xt}-4\nu+4\mathrm{i}\mu\nu\nu_{x}=0,
\end{align}
\end{subequations}
which, under the coordinate scaling $(t, x)\rightarrow(t/2, -x/2)$, is precisely the Mikhailov model \eqref{MM}.
The plane-wave factors in the continuous limit read
\begin{subequations}
\label{ctn-plane-def}
\begin{align}
&\rho_{\iota}(z)=e^{(-1)^{\iota-1}(zt+z^{-1} x)}\rho^{0}_{\iota}(z),\\
&\sigma_{\iota}(z)=e^{(-1)^{\iota}(zt+z^{-1} x)}\sigma^{0}_{\iota}(z).
\end{align}
\end{subequations}


\section{Local and nonlocal reductions}\label{Sec-Redu}

The four families of conjugate-symmetric discrete Mikhailov models constructed in Section 2 are invariant under the specified conjugation transformation, which allows for
consistent complex-conjugation reductions.
In this section, we perform both local and nonlocal symmetry reductions on the symmetric discrete Mikhailov models. For local reductions, we directly 
impose the conjugate constraint $\nu=\mu^*$ on each family of symmetric models to obtain four families of local dFL equations. 
For nonlocal reductions, we first recombine the lattice equations of the four symmetric families in a pairwise manner to obtain suitable combined systems, and 
then apply the nonlocal constraint $\nu=\mathrm{i}\mu_{-}^*$ to derive 
four families of nonlocal dFL equations. For both classes of dFL type equations, we construct their exact soliton and multiple-pole solutions via the Cauchy 
matrix approach, and present a systematic analysis of their typical dynamical properties.

\subsection{Local reduction}
For the  symmetric discrete Mikhailov model \eqref{dFL1}, we observe that setting $\nu=\mu^*$ reduces the two equations in the model 
to a single equation, which we refer to as the dFL equation. This reduction can also be realized at the solution level by imposing 
appropriate constraints. Models \eqref{dFL2}, \eqref{dFL3}, and \eqref{dFL4} remain valid under this reduction. As a representative example, 
the reduction of model \eqref{dFL1} gives
\begin{align}
\label{eq-dFL}
(\wh{\mu}+\wt{\mu})\big((\mu^*+\wh{\wt{\mu}}^*)(\wh{\mu}+\wh{\wt{\mu}})-\mathrm{i}\wh{P}_{1}-\mathrm{i}Q_{1}\big)
-(\mu+\wh{\wt{\mu}})\big((\wh{\mu}^*+\wt{\mu}^*)(\mu+\wt{\mu})+\mathrm{i}Q_{1}-\mathrm{i}P_{1}\big)=0,
\end{align}
where
\begin{align}
\label{P1Q1}
P_{1}=(4p^2-|\mu+\wt{\mu}|^4)^{\frac{1}{2}},\quad Q_{1}=(4/q^2-|\mu-\wh{\mu}|^4)^{\frac{1}{2}}.
\end{align}
The remaining three dFL equations, obtained from models \eqref{dFL2}-\eqref{dFL4} under the same reduction, are listed in Appendix B. 
\begin{remark}
Under the reduction $\nu = \mu^*$, for the two equations in each model to reduce to a single equation (i.e., they become mutually conjugate), the associated variables $P_{1}$ and $Q_{1}$ must satisfy 
\begin{align}
\label{PQ-Real}
P_{1}=P_{1}^*,\quad Q_{1}=Q_{1}^*.
\end{align}
We note that \eqref{PQ-Real} holds trivially when the lattice parameters $p$ and $q$ are sufficiently large. We can expand the radicand of $P_{1}$ following the continuous limit approach presented in Section 4, to obtain
\begin{align}
4p^2-|\mu+\wt{\mu}|^4=4p^2-|2\mu+\mu_{t}/p+\mu_{tt}/(2p^2)+\cdots|^4=4p^2-\mathcal{O}(1).
\end{align}
Hence, when $p$ is sufficiently large, $4p^2$ dominates and the expression remains strictly positive, which implies that $P_{1}$ is positive. Likewise, expanding the radicand of $Q_{1}$ gives
\begin{align}
4/q^2-|\mu_{x}/q+\mu_{xx}/(2q^2)+\cdots|^4=4/q^2-\mathcal{O}(q^{-4}),
\end{align}
which implies that $Q_{1}$ is also strictly positive for sufficiently large $q$.
\end{remark}

We demonstrate the validity of this reduction at the solution
level in the following theorem.
\begin{theorem}\label{Theo-AKNS-redu}
The AKNS-type solution \eqref{sym-FL-solu} admits local reduction $\nu=\mu^{*}$ under constraints
		\begin{align}
			\label{AKNS-redu1}
			\bK_{2}=-\bK_{1}^{*},\quad \br_{2}=\mathrm{i}\bK_{2}\br_{1}^{*},\quad \bs_{2}=\bs_{1}^{*}.
		\end{align}
	\end{theorem}
	\begin{proof}
		By condition \eqref{AKNS-redu1}, from \eqref{DES-sym-ent} one immediately obtains
		\begin{align}
			\bK_{1}\bM_{1}+\bM_{1}\bK_{1}^{*}=\br_{1}(\bs_{1}^{\st})^{*},\quad
			\bK_{1}(\mathrm{i}\bK_{1}^{-1}\bM_{2}^{*})+(\mathrm{i}\bK_{1}^{-1}\bM_{2}^{*})\bK_{1}^{*}=\br_{1}(\bs_{1}^{\st})^{*}.
		\end{align}
		Subtracting the second equation from the first one and using the Sylvester Theorem, 
		one easily finds the relations between $\bM_{1}$ and $\bM_{2}$: 
		\begin{align}
			\bM_{2}^{*}=-\mathrm{i}\bK_{1}\bM_{1},\quad\bM_{1}^{*}=\mathrm{i}\bK_{2}^{-1}\bM_{2}\quad\Longrightarrow\quad\bM_{1}^{*}\bM_{2}^{*}=\bK_{2}^{-1}\bM_{2}\bK_{1}\bM_{1}.
		\end{align}
		Through straightforward calculations, it is demonstrated that
		\begin{align*}
			u_{3}^{*}v_{4}=&(\bs_{1}^{\st})^{*}(\bI-\bM_{1}^{*}\bM_{2}^{*})^{-1}\br_{1}^{*}\left[1+\bs_{1}^{\st}\bM_{1}(\bI-\bM_{2}\bM_{1})^{-1}\bK_{2}^{-1}\br_{2}\right] \\
			=&-\mathrm{i}\bs_{2}^{\st}(\bI-\bM_{1}^{*}\bM_{2}^{*})^{-1}(\bI-\bM_{2}\bM_{1}+\bK_{2}^{-1}\br_{2}\bs_{1}^{\st}\bM_{1})(\bI-\bM_{2}\bM_{1})^{-1}\bK_{2}^{-1}\br_{2}\\
			=&-\mathrm{i}\bs_{2}^{\st}(\bI-\bM_{1}^{*}\bM_{2}^{*})^{-1}(\bI-\bK_{2}^{-1}\bM_{2}\bK_{1}\bM_{1})(\bI-\bM_{2}\bM_{1})^{-1}\bK_{2}^{-1}\br_{2}\\
			=&-\mathrm{i}\bs_{2}^{\st}(\bI-\bM_{2}\bM_{1})^{-1}\bK_{2}^{-1}\br_{2}=\mathrm{i}v_{2},
		\end{align*}
where the equation \eqref{sym-SE-12} is applied. Therefore, the proof is completed.
\end{proof}
	
The following theorem reveals the AKNS-type Cauchy matrix soliton solution 
and multiple-pole solution to the dFL equation \eqref{eq-dFL}.
\begin{theorem}\label{Theo-AKNS-solu}
The function 
\begin{align}
\label{AKNS-FLeq-solu}
\mu=\bs_{1}^{\st}(\bI-\mathrm{i}\bM_{1}\bK^{*}_{1}\bM^{*}_{1})^{-1}\br_{1},
\end{align}
provides i) the AKNS-type Cauchy matrix $\mathcal{N}$-soliton solution for the dFL equation \eqref{eq-dFL} with
		\begin{subequations}
			\label{AKNS-msr-def}
			\begin{align}
				\label{AKNS-k-def}
				&\bK_{1}=\mathrm{diag}(k_{1},k_{2},\cdots,k_{\mathcal{N}}),\quad k_{i}\in\mathbb{C}, \quad i=1,2,\cdots,\mathcal{N},\\
				\label{AKNS-r-def}
				&\ts_{1}=(\sigma_{2}(k_{1}),\sigma_{2}(k_{2}),\cdots,\sigma_{2}(k_{\mathcal{N}})),\quad 
				\br_{1}=(\rho_{1}(k_{1}),\rho_{1}(k_{2}),\cdots,\rho_{1}(k_{\mathcal{N}}))^{\st},\\
				\label{AKNS-m1-def}
				&\bM_{1}=(M_{1,ij})_{\mathcal{N}\times \mathcal{N}},\quad M_{1,ij}=\frac{\rho_{1}(k_{i})\sigma_{2}^{*}(k_{j})}{k_{i}+k_{j}^{*}};
			\end{align}
		\end{subequations}
		and ii) the AKNS-type Cauchy matrix $\mathcal{N}$-pole solution for the dFL equation \eqref{eq-dFL} with
		\begin{subequations}
			\label{AKNS-msr-def-1}
			\begin{align}
				&\bK_{1}=\bJ[k,1],\quad\br_{1}=\bF_{k}[\rho_{1}(k)]\bar{\be}_{N},\quad\ts_{1}=\bar{\be}_{N}^{\st}\bH_{k}[\sigma_{2}(k)],\\
				&\bM_{1}=\bF_{k}[\rho_{1}(k)]\bar{\bG}\bH^{*}_{k}[\sigma_{2}(k)], \quad
				\bar{\bG}=(\bar{G}_{i,j})_{\mathcal{N}\times \mathcal{N}},\quad \bar{G}_{i,j}=\frac{(-1)^{i+j}\mathrm{C}^{i-1}_{i+j-2}}{(k+k^{*})^{i+j-1}}.
			\end{align}
\end{subequations}
Here $k_{i}+k_{j}^*\neq0$. The discrete plane-wave factors $\rho_{1}(z)$ and $\sigma_{2}(z)$ are defined in \eqref{plane-def}, and the matrices $\bJ,\bF,\bH$ are defined in \eqref{J-def} and \eqref{FH-def}, respectively.
\end{theorem}
	
In what follows, we focus on the explicit expressions of the one-, two-soliton
solution and the 2-pole solution, as well as their dynamical behaviors.
	
$\bullet$\hspace{1em}\textit{One-soliton solution}. We take $\mathcal{N}=1$ in \eqref{AKNS-FLeq-solu} and \eqref{AKNS-msr-def}. 
Then the one-soliton solution can be readily expressed as
\begin{align}
\label{1-soliton-AKNS}
\mu=\frac{(k_{1}+k_{1}^{*})^{2}\mathcal{A}(k_{1})}{(k_{1}+k_{1}^{*})^{2}-\mathrm{i}k_{1}^{*}\abs{\mathcal{A}(k_{1})}^{2}},
\end{align}
where function 
\begin{align}
	\label{A-def}
		\mathcal{A}(z)=\rho_{1}(z)\sigma_{2}(z)=\left( \frac{p+z}{p-z}\right)^{n}
		\left(\frac{q+z^{-1}}{q-z^{-1}}\right)^{m}\mathcal{A}^{0}(z),
\end{align}
and $\mathcal{A}^{0}(z)$ is a constant independent of $(n,m)$.
	
To understand the dynamical behaviors of this solution, we write down its envelope
\begin{align}
		\label{1-soliton-AKNS-modulus}
		\abs{\mu}^{2}=\frac{1}{
			\frac{\abs{k_{1}}^{2}\abs{\mathcal{A}(k_{1})}^{2}}{16\Re(k_{1})^{4}}+\frac{1}{\abs{\mathcal{A}(k_{1})}^{2}}-\frac{\Im(k_{1})}{2\Re(k_{1})^{2}}},
\end{align}
where $\text{Re}(k)$ and $\text{Im}(k)$ represent real and imaginary parts of $k$, respectively.
This solution describes a stable travelling wave that is single peaked and unidirectional.
Its height relative to $|\mu|^2=0$ is $2\Re(k_{1})^2/(\abs{k_{1}}-\Im(k_{1}))$, the velocity and peak trajectory 
are given by $-\alpha_{1}/\beta_{1}$, and
$n\alpha_{1}+m\beta_{1}+\gamma_{1}=0$, respectively, where
\begin{align}
\label{alpha-beta-def}
\alpha_{i}\doteq\ln\abs{\frac{p+k_{i}}{p-k_{i}}},\quad\beta_{i}\doteq\ln\abs{\frac{q+k^{-1}_{i}}{q-k^{-1}_{i}}},\quad\gamma_{i}\doteq
\ln\frac{\abs{\mathcal{A}^{0}_i}\sqrt{\abs{k_{i}}}}{2\abs{\Re(k_{i})}},\quad i=1,2.
\end{align}
We depict solution \eqref{1-soliton-AKNS-modulus} in Figure 1. 
\vskip35pt
\begin{center}
\begin{picture}(120,80)
\put(-150,-23){\resizebox{!}{4.0cm}{\includegraphics{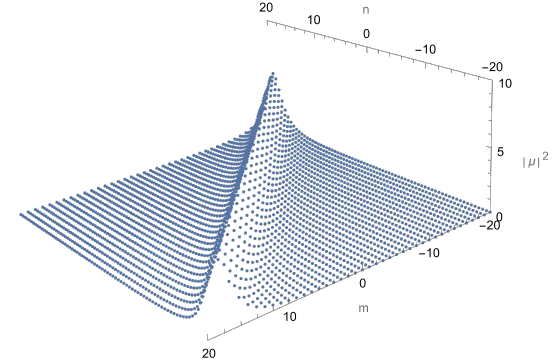}}}
\put(80,-23){\resizebox{!}{4cm}{\includegraphics{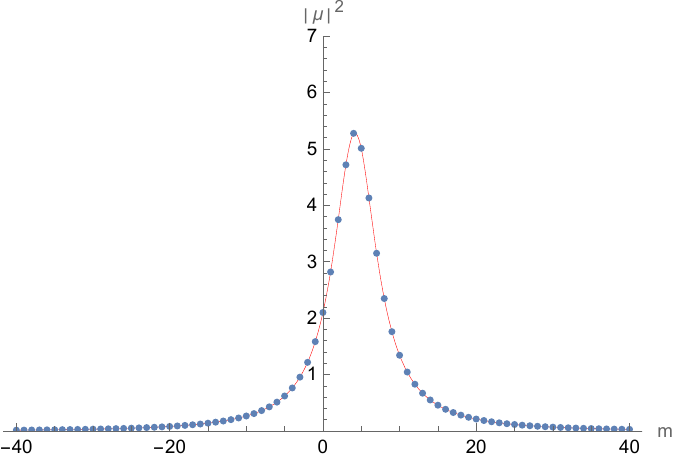}}}
\end{picture}
\end{center}
\vskip 20pt
\begin{center}
\begin{minipage}{15cm}{\footnotesize
\qquad\qquad\qquad\qquad\quad(a)\qquad\qquad\qquad\qquad\qquad\qquad\qquad\qquad\qquad\qquad\qquad\quad (b) \\
{\bf Figure 1.} One-soliton solution $|\mu|^2$ given by \eqref{1-soliton-AKNS-modulus} with 
$k_{1}=-0.35 + 1.3\mathrm{i},~p=q=10$ and $\mathcal{A}^{0}(k_1)=1$:
(a) shape and movement; (b) 2D-plot of (a) at $n=5$.}
\end{minipage}
\end{center}
	
$\bullet$\hspace{1em}\textit{Two-soliton solution}. Now, we take $\mathcal{N}=2$ in \eqref{AKNS-FLeq-solu} and \eqref{AKNS-msr-def}.  
After a series of calculations, the two-soliton solution is expressed as 
\begin{align}
\label{2-soliton-AKNS-mu}
\mu=g/f
\end{align}
with
\begin{subequations}
\label{2-soliton-AKNS-fg}
\begin{align}
\label{f-def-1}
&f=\abs{\bI-\mathrm{i}\bM_{1}\bK^{*}_{1}\bM^{*}_{1}}=1-\mathrm{i}\sum_{i=1}^{2}\sum_{j=1}^{2}
\frac{k_{i}^{*}\mathcal{A}^{*}(k_{i})\mathcal{A}(k_{j})}{(k_{i}^{*}+k_{j})^{2}}-\frac{\abs{k_{1}-k_{2}}^{4}
\prod_{i=1}^{2}k_{i}^{*}|\mathcal{A}(k_{i})|^2}{\prod_{i=1}^{2}\prod_{j=1}^{2}(k_{i}+k_{j}^{*})^{2}},\\
\label{g-def-1}
&g=\mathcal{A}(k_{1})+\mathcal{A}(k_{2})-\mathrm{i}(k_{1}-k_{2})^{2}\mathcal{A}(k_{1})\mathcal{A}(k_{2})
\sum_{i=1}^{2}\frac{k_{i}^{*}\mathcal{A}^{*}(k_{i})}{(k_{i}+k_{i}^{*})^{2}(k_{3-i}+k_{i}^{*})^{2}}.
\end{align}
\end{subequations}
	
In the following we will  examine the asymptotic form of the 2-soliton solution \eqref{2-soliton-AKNS-mu} 
as $n\rightarrow \pm\infty$ by means of the moving coordinate expansions. To facilitate the analysis, we define 
\begin{align}
\label{xi-def}
\xi_{i}=\text{Ln}\mathcal{A}(k_{i}),\quad i=1,2,
\end{align}
where the notation $\text{Ln} z$ denotes a chosen branch of the complex logarithm, 
i.e., $\text{Ln} z=\ln\abs{z}+\mathrm{i}\text{Arg}z$. A direct result is that
\begin{align}
\Re(\xi_{i})=\ln\abs{\mathcal{A}(k_{i})}=n\alpha_{i}+m\beta_{i}+\xi_{i}^{0},\quad i=1,2,
\end{align}
where $\xi_{i}^{0}\doteq\ln\abs{\mathcal{A}^{0}(k_i)}$, and $\alpha_{i},\beta_{i}$ are defined in \eqref{alpha-beta-def}.
For convenience, we call the asymptotic solitons as $k_1$-soliton and $k_2$-soliton, respectively, and we
impose $0<\alpha_1<\beta_2<\beta_1<\alpha_2$ to guarantee $\vartheta=\alpha_{2}\beta_1-\alpha_{1}\beta_2>0$.
	
First, we hold $\Re(\xi_{1})$ fixed, and asymptotically expand $f$ and $g$ for
large $n$ with 
\begin{align}
\Re(\xi_{2})=\frac{\beta_{2}}{\beta_{1}}\Re(\xi_{1})+\frac{\vartheta}{\beta_1} n+\beta_{2}\left(\frac{\xi_{2}^{0}}{\beta_{2}}-\frac{\xi_{1}^{0}}{\beta_{1}}\right).
\end{align}
This yields, after neglecting subdominant terms,
\begin{align}
\label{sy-2ss-f-asymp-k1}
f\sim\left\{
\begin{aligned}
& -\frac{\mathrm{i}k_{2}^{*}\abs{\mathcal{A}(k_{2})}^{2}}{(k_{2}+k_{2}^{*})^{2}}-\frac{\abs{k_{1}-k_{2}}^{4}
\prod_{i=1}^{2}k_{i}^{*}|\mathcal{A}(k_{i})|^2}{\prod_{i=1}^{2}\prod_{j=1}^{2}(k_{i}+k_{j}^{*})^{2}}, \quad n\rightarrow +\infty, \\
& 1-\frac{\mathrm{i}k_{1}^{*}|\mathcal{A}(k_{1})|^2}{(k_{1}+k_{1}^{*})^{2}},\quad n\rightarrow -\infty,
\end{aligned}	
\right.
\end{align}
\begin{align}
\label{sy-2ss-g-asymp-k1}
g\sim\left\{
\begin{aligned}
& -\frac{\mathrm{i}(k_{1}-k_{2})^{2}k_{2}^{*}\mathcal{A}(k_{1})\abs{\mathcal{A}(k_{2})}^{2}}{(k_{1}+k_{2}^{*})^{2}(k_{2}+k_{2}^{*})^{2}}, 
\quad n\rightarrow +\infty, \\
& \mathcal{A}(k_{1}),\quad n\rightarrow -\infty,
\end{aligned}	
\right.
\end{align}
and hence we obtain the envelope of $k_1$-soliton:
\begin{align}
		\label{sy-2ss-mu-asymp-k1}
		|\mu|^2\sim\left\{
		\begin{aligned}
			& \frac{1}{\frac{\tau^2\abs{k_{1}}^{2}\abs{\mathcal{A}(k_{1})}^{2}}{16\Re(k_{1})^{4}}
				+\frac{1}{\tau^2\abs{\mathcal{A}(k_{1})}^{2}}-\frac{\Im(k_{1})}{2\Re(k_{1})^{2}}}, 
			\quad n\rightarrow +\infty, \\
			& \frac{1}{
				\frac{\abs{k_{1}}^{2}\abs{\mathcal{A}(k_{1})}^{2}}{16\Re(k_{1})^{4}}+\frac{1}{\abs{\mathcal{A}(k_{1})}^{2}}
				-\frac{\Im(k_{1})}{2\Re(k_{1})^{2}}}, \quad n\rightarrow -\infty,
		\end{aligned}	
		\right.
	\end{align}
	where
	\begin{align}
		\tau\doteq \frac{(\Re( k_{1}-k_{2}))^{2}+(\Im(k_{1}-k_{2}))^{2}}{(\Re(k_{1}+k_{2}))^{2}+(\Im(k_{1}-k_{2}))^{2}}.
	\end{align}
In the asymptotic past, the $k_1$-soliton \eqref{sy-2ss-mu-asymp-k1} has peak intensity $2\Re(k_{1})^2/(\abs{k_{1}}-\Im(k_{1}))$ and propagates with
velocity $-\alpha_{1} /\beta_{1}$, peak trajectory $n\alpha_{1}+m\beta_{1}+\gamma_{1}=0$.
Similarly, in the asymptotic future, the $k_1$-soliton \eqref{sy-2ss-mu-asymp-k1} retains its amplitude and velocity,
whereas the peak trajectory is given by $n\alpha_{1}+m\beta_{1}+\gamma_{1}+\ln\tau=0$. Naturally, the phase shift is determined by $-\ln \tau/\beta_1$.
	
Second, we hold $\text{Re}(\xi_2)$ fixed, and asymptotically expand $f$ and $g$ for large $n$ with 
	\begin{align}
		\Re(\xi_{1})=\frac{\beta_{1}}{\beta_{2}}\Re(\xi_{2})-\frac{1}{\beta_{2}}\vartheta n+\beta_{1}
		\left( \frac{\xi_{1}^{0}}{\beta_{1}}-\frac{\xi_{2}^{0}}{\beta_{2}}\right).
	\end{align}
After neglecting subdominant terms, we obtain
	\begin{align}
		\label{sy-2ss-f-asymp-k2}
		f\sim\left\{
		\begin{aligned}
			& 1-\frac{\mathrm{i}k_{2}^{*}|\mathcal{A}(k_{2})|^{2}}{(k_{2}+k_{2}^{*})^{2}}, \quad n\rightarrow +\infty, \\
			& -\frac{\mathrm{i}k_{1}^{*}\abs{\mathcal{A}(k_{1})}^{2}}{(k_{1}+k_{1}^{*})^{2}}-\frac{\abs{k_{1}-k_{2}}^{4}
				\prod_{i=1}^{2}k_{i}^{*}|\mathcal{A}(k_{i})|^2}{\prod_{i=1}^{2}\prod_{j=1}^{2}(k_{i}+k_{j}^{*})^{2}},\quad n\rightarrow -\infty,
		\end{aligned}	
		\right.
	\end{align}
	\begin{align}
		\label{sy-2ss-g-asymp-k2}
		g\sim\left\{
		\begin{aligned}
			& \mathcal{A}(k_{2}), 
			\quad n\rightarrow +\infty, \\
			& -\frac{\mathrm{i}(k_{2}-k_{1})^{2}k_{1}^{*}\mathcal{A}(k_{2})
				\abs{\mathcal{A}(k_{1})}^{2}}{(k_{2}+k_{1}^{*})^{2}(k_{1}+k_{1}^{*})^{2}},\quad n\rightarrow -\infty,
		\end{aligned}	
		\right.
	\end{align}
which yields the envelope of $k_2$-soliton
	\begin{align}
		\label{sy-2ss-mu-asymp-k2}
		|\mu|^2\sim\left\{
		\begin{aligned}
			&  \frac{1}{
				\frac{\abs{k_{2}}^{2}\abs{\mathcal{A}(k_{2})}^{2}}{16\Re(k_{2})^{4}}+\frac{1}{\abs{\mathcal{A}(k_{2})}^{2}}-\frac{\Im(k_{2})}{2\Re(k_{2})^{2}}}, 
			\quad n\rightarrow +\infty, \\
			&  \frac{1}{
				\frac{\tau^2\abs{k_{2}}^{2}\abs{\mathcal{A}(k_{2})}^{2}}{16\Re(k_{2})^{4}}+
				\frac{1}{\tau^2\abs{\mathcal{A}(k_{2})}^{2}}-\frac{\Im(k_{2})}{2\Re(k_{2})^{2}}}, \quad n\rightarrow -\infty.
		\end{aligned}	
		\right.
	\end{align}
In the asymptotic past, the $k_2$-soliton \eqref{sy-2ss-mu-asymp-k2} travels with
amplitude $2\Re(k_{2})^2/(\abs{k_{2}}-\Im(k_{2}))$, velocity $-\alpha_{2} /\beta_{2}$, peak trajectory $n\alpha_{2}+m\beta_{2}+\gamma_{2}+\ln\tau=0$.
In the asymptotic future, the $k_2$-soliton \eqref{sy-2ss-mu-asymp-k2} retains its amplitude and velocity,
whereas the peak trajectory is given by $n\alpha_{2}+m\beta_{2}+\gamma_{2}=0$. Thus, the phase shift is determined by $\ln \tau/\beta_2$.
	
The interaction of two-soliton solution is shown in Figure 2.
	\vskip35pt
	\begin{center}
		\begin{picture}(120,80)
			\put(-160,-23){\resizebox{!}{3.5cm}{\includegraphics{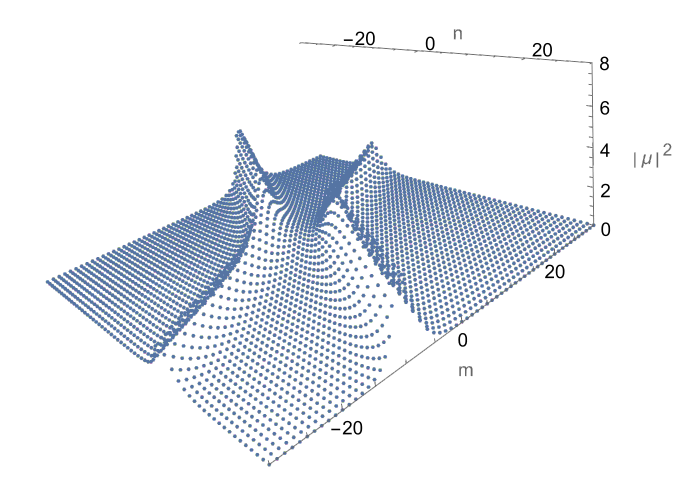}}}
			\put(-10,-23){\resizebox{!}{3.5cm}{\includegraphics{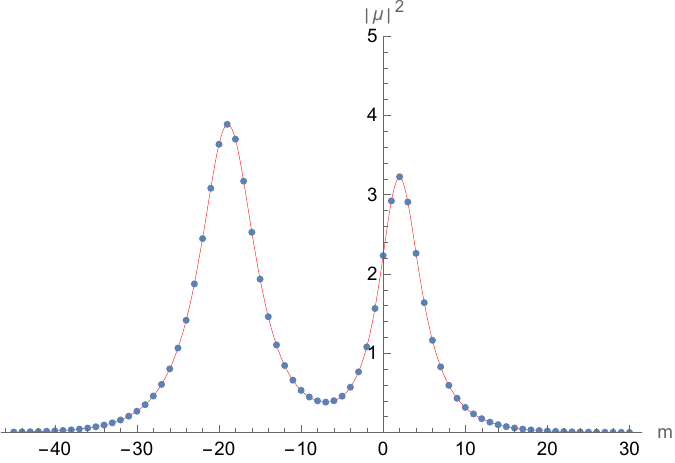}}}
			\put(140,-23){\resizebox{!}{3.5cm}{\includegraphics{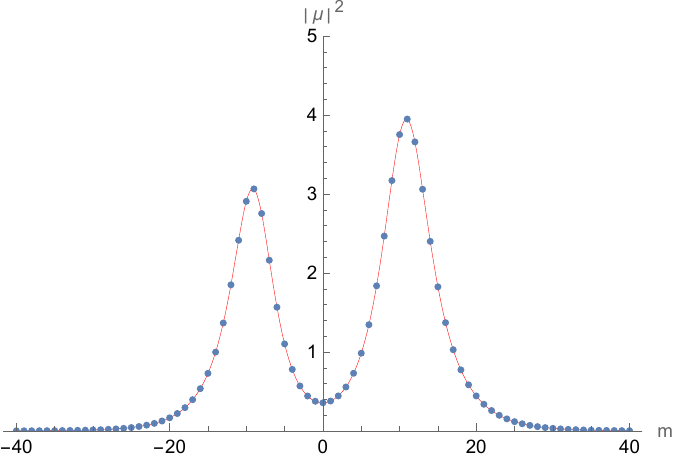}}}
		\end{picture}
	\end{center}
	\vskip 20pt
	\begin{center}
		\begin{minipage}{15cm}{\footnotesize
				\qquad\qquad\qquad(a)\qquad\qquad\qquad\qquad\qquad\qquad\qquad\qquad (b) \qquad\qquad\qquad\qquad\qquad\qquad\qquad (c)\\
				{\bf Figure 2.} Two-soliton solution $|\mu|^2$ given by \eqref{2-soliton-AKNS-mu} with 
				$k_{1}=-1.3 + 0.6\mathrm{i},~k_{2}=0.7 + 0.2\mathrm{i},~p=q=10$ and $\mathcal{A}^{0}(k_1)=\mathcal{A}^{0}(k_2)=1$:
				(a) shape and movement; (b) 2D-plot of (a) at $n=6$; (c) 2D-plot of (a) at $n=0$.}
		\end{minipage}
	\end{center}
	
$\bullet$\hspace{1em}\textit{The 2-pole solution}. To obtain the 2-pole solution, we take 
$\mathcal{N}=2$ in \eqref{AKNS-FLeq-solu} and \eqref{AKNS-msr-def-1} and assume that constant $\mathcal{A}^{0}(k)$ is independent of $k$. Then we have $\mu=g/f$, where
\begin{subequations}
		\label{lr-sy-2pl}
		\begin{align}
			f=&(k+k^{*})^{8}-k^{*^2}\abs{\mathcal{A}(k)}^{4}-\mathrm{i}(k+k^{*})^{4}\abs{\mathcal{A}(k)}^{2}\big[6k^{*}
			-2(k+k^{*})\big(1+k^*(\eta+\eta^*)\big) \nn \\
			&+(k+k^{*})^{2}(\eta+k^*\abs{\eta}^2)\big] ,\\
			g=&(k+k^{*})^{3}\mathcal{A}(k)\big[\eta(k+k^{*})^{5}+\mathrm{i}\abs{\mathcal{A}(k)}^{2}\big(\eta^*k^*(k+k^*)+(k-3k^*)\big)\big], 
		\end{align}
	\end{subequations}
where $\eta=2\left(\frac{np}{p^2-k^2}+\frac{mq}{1-k^2q^2}\right)$. We depict $|\mu|^2$ for this solution in Figure 3.
	
	\vskip35pt
	\begin{center}
		\begin{picture}(120,80)
			\put(-160,-23){\resizebox{!}{3.5cm}{\includegraphics{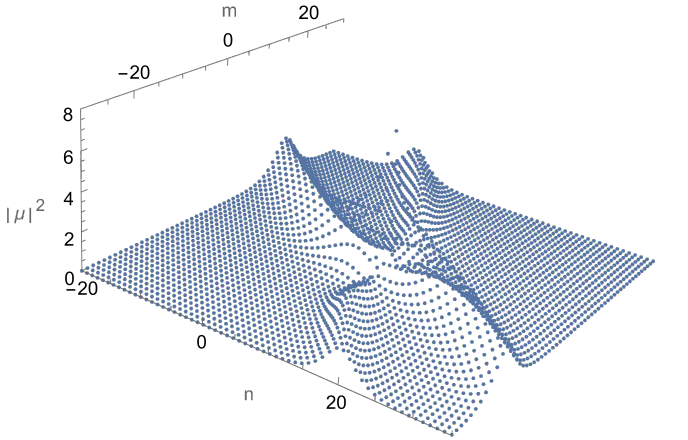}}}
			\put(-10,-23){\resizebox{!}{3.5cm}{\includegraphics{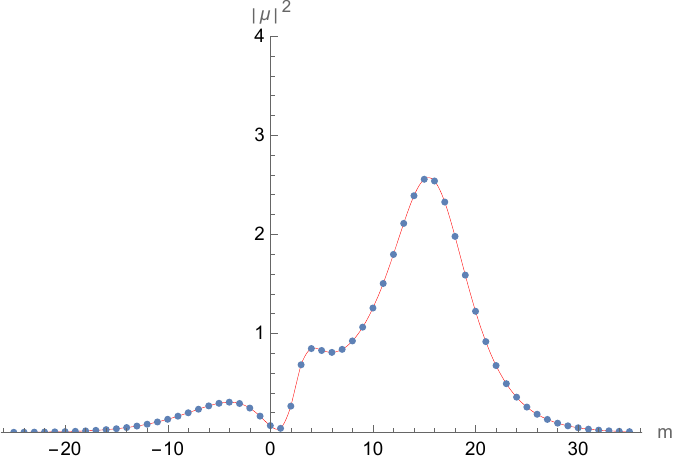}}}
			\put(140,-23){\resizebox{!}{3.5cm}{\includegraphics{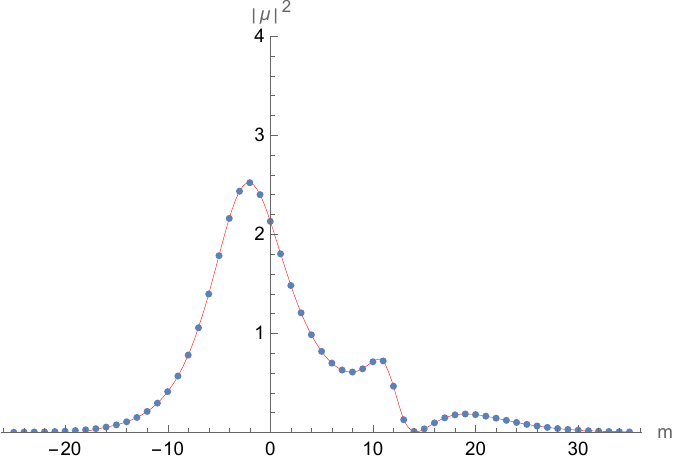}}}
		\end{picture}
	\end{center}
	\vskip 20pt
\begin{center}
		\begin{minipage}{15cm}{\footnotesize
				\qquad\qquad\qquad\qquad(a)\qquad\qquad\qquad\qquad\qquad\qquad\quad (b) \qquad\qquad\qquad\qquad\qquad\qquad\qquad (c)\\
				{\bf Figure 3.} 2-pole solution $|\mu|^2$ given by \eqref{lr-sy-2pl} with 
				$k=1,~p=q=10$ and $\mathcal{A}^{0}(k)=1$:
				(a) shape and movement; (b) 2D-plot of (a) at $n=1$; (c) 2D-plot of (a) at $n=9$.}
		\end{minipage}
\end{center}
	
\subsection{Nonlocal reduction}
We recombine the lattice equations in models \eqref{dFL1}-\eqref{dFL4} pairwise so that the resulting four new models admit nonlocal reductions. 
After careful consideration, we pair \eqref{dFL1-1} with \eqref{dFL4-2}, \eqref{dFL2-1} with \eqref{dFL3-2}, \eqref{dFL3-1} with \eqref{dFL2-2}, and \eqref{dFL4-1} 
with \eqref{dFL1-2}, thereby obtaining four recombined discrete Mikhailov models, denoted by $(N.1)$-$(N.4)$. These models themselves admit nonlocal
reduction $\nu=\mathrm{i}\mu_{-}^*$, then from $(N.1)$-$(N.4)$ one gets the four distinct nonlocal dFL equations. As a representative example, the reduction of model $(N.1)$ 
gives\footnote{For the notation $f_{-}=f(-n,-m)$, the shifts of the
reflected field are defined by
$\wt{f}_{-}=f(-n-1,-m)$, $\wh{f}_{-}=f(-n,-m-1)$
and $\wh{\wt{f}}_{-}=f(-n-1,-m-1)$.}
\begin{align}
\label{non-eqFL}
(\wh{\mu}+\wt{\mu})\big((\mu_{-}^*&+\wh{\wt{\mu}}_{-}^*)(\wh{\mu}+\wh{\wt{\mu}})-\wh{P}_{2}-Q_{2}\big)
-(\mu+\wh{\wt{\mu}})\big((\wh{\mu}_{-}^*+\wt{\mu}_{-}^*)(\mu+\wt{\mu})+Q_{2}-P_{2}\big)=0,
\end{align}
where
\begin{align}
\label{P2Q2}
P_{2}=\big(4p^2+(\mu+\wt{\mu})^2(\mu_{-}^*+\wt{\mu}_{-}^*)^2\big)^{\frac{1}{2}},\quad Q_{2}=\big(4/q^2+(\mu-\wh{\mu})^2(\mu_{-}^*-\wh{\mu}_{-}^*)^2\big)^{\frac{1}{2}}.
\end{align}
The remaining three nonlocal dFL equations are listed in Appendix C.
	
Different from the local dFL equation \eqref{eq-dFL}, 
the nonlocal dFL equation \eqref{non-eqFL} has both AKNS-type and KP-type solutions. In other words, the AKNS-type solution \eqref{sym-FL-solu} and 
the KP-type solution \eqref{asym-FL-solu} admit nonlocal reduction $\nu=\mathrm{i}\mu^{*}_{-}$.
	
\subsubsection{AKNS-type solution}
	
The AKNS-type solution \eqref{sym-FL-solu} admits nonlocal reduction $\nu=\mathrm{i}\mu^{*}_{-}$, which is summarized in the following theorem.
\begin{theorem}\label{Th-AK-non}
The AKNS-type solution \eqref{sym-FL-solu} admits $\nu=\mathrm{i}\mu^{*}_{-}$ under the constraints
\begin{align}
\label{AKNS-nonlocal-redu1}
\bK_{2}=\bK_{1}^{\dagger}, \quad \br_{2}=\bK_{2}(\bs^{\st}_{1,-})^{\dagger}, \quad \bs^{\st}_{2}=-\br_{1,-}^{\dagger},
\end{align}
where notation $\dagger$ represents the Hermitian adjoint.
\end{theorem}
\begin{proof}
By \eqref{AKNS-nonlocal-redu1}, we observe that the Sylvester equations \eqref{DES-sym-ent} become
\begin{align}
\label{nl-sym-Sy}		
\bK_{1}\bM_{1}-\bM_{1}\bK_{1}^{\dagger}=-\br_{1}\br_{1,-}^{\dagger},\quad\bK_{1}^{\dagger}\bM_{2}-\bM_{2}\bK_{1}
=\bK_{1}^{\dagger}(\bs^{\st}_{1,-})^{\dagger}\bs^{\st}_{1}.
\end{align} 
Taking the Hermitian conjugate of the two above equations yields
\begin{align}
\label{nl-sym-Sy-1}	
\bM^{\dagger}_{1,-}\bK_{1}^{\dagger}-\bK_{1}\bM^{\dagger}_{1,-}=-\br_{1}\br_{1,-}^{\dagger},\quad
\bM^{\dagger}_{2,-}\bK_{1}-\bK_{1}^{\dagger}\bM^{\dagger}_{2,-}=(\bs^{\st}_{1,-})^{\dagger}\bs^{\st}_{1}\bK_{1}.
\end{align}
Comparing \eqref{nl-sym-Sy} and \eqref{nl-sym-Sy-1}	and noticing the Sylvester equation theorem, we find
\begin{align} 
\bM_{2,-}^{\dagger}=-(\bK_{1}^{\dagger})^{-1}\bM_{2}\bK_{1},\quad\bM_{1,-}^{\dagger}
			=-\bM_{1}\quad\Longrightarrow\quad\bM_{2,-}^{\dagger}\bM_{1,-}^{\dagger}=\bK_{2}^{-1}\bM_{2}\bK_{1}\bM_{1}.
\end{align}
Following a series of computations, we arrive at 
\begin{align*}
u_{3,-}^{*}v_{4}=&u_{3,-}^{\dagger}v_{4}\\
=&\br_{1,-}^{\dagger}\left(\bI-\bM_{2,-}^{\dagger}\bM_{1,-}^{\dagger}\right)^{-1}(\bs^{\st}_{1,-})^{\dagger}
\left(1+\bs^{\st}_{1}\bM_{1}(\bI-\bM_{2}\bM_{1})^{-1}\bK_{2}^{-1}\br_{2}\right) \\
=&-\bs^{\st}_{2}\left(\bI-\bM_{2,-}^{\dagger}\bM_{1,-}^{\dagger}\right)^{-1}
(\bI-\bM_{2}\bM_{1}+\bK_{2}^{-1}\br_{2}\bs^{\st}_{1}\bM_{1})(\bI-\bM_{2}\bM_{1})^{-1}\bK_{2}^{-1}\br_{2}\\
=&-\bs^{\st}_{2}\left( \bI-\bM_{2,-}^{\dagger}\bM_{1,-}^{\dagger}\right)^{-1}
(\bI-\bK_{2}^{-1}\bM_{2}\bK_{1}\bM_{1})(\bI-\bM_{2}\bM_{1})^{-1}\bK_{2}^{-1}\br_{2}\\
=&-\bs^{\st}_{2}(\bI-\bM_{2}\bM_{1})^{-1}\bK_{2}^{-1}\br_{2}=v_{2}.
\end{align*}
The verification of this theorem is thus completed.
\end{proof}
	
We present the AKNS-type Cauchy matrix soliton solution and multiple-pole solution to the nonlocal dFL equation \eqref{non-eqFL} in the following theorem. 
\begin{theorem}\label{Theo-AKNS-solu-non}
The function 
\begin{align}
\label{AKNS-nonlocal-FLeq-solu} 
\mu=\bs^{\st}_{1}(\bI-\bM_{1}\bM_{2})^{-1}\br_{1},
\end{align}
provides i) the AKNS-type Cauchy matrix $\mathcal{N}$-soliton solution for the nonlocal dFL equation \eqref{non-eqFL} with
\begin{subequations}
\label{AKNS-nonlocal-msr-def}
\begin{align}
\label{AKNS-non-k-def}
&\bK_{1}=\mathrm{diag}(k_{1},k_{2},\cdots,k_{\mathcal{N}}),\quad k_{i}\in\mathbb{C},\quad i=1,2,\cdots,\mathcal{N},\\
\label{AKNS-non-r-def}
&\ts_{1}=(\sigma_{2}(k_{1}),\sigma_{2}(k_{2}),\cdots,\sigma_{2}(k_{\mathcal{N}})),\quad\br_{1}=(\rho_{1}(k_{1}),\rho_{1}(k_{2}),\cdots,\rho_{1}(k_{\mathcal{N}}))^{\st},\\
\label{AKNS-non-m1-def}
&\bM_{1}=(M_{1,ij})_{\mathcal{N}\times \mathcal{N}},\quad M_{1,ij}=\frac{-\rho_{1}(k_{i})\rho_{1,-}^{*}(k_{j})}{k_{i}-k_{j}^{*}},\\
\label{AKNS-non-m2-def}
&\bM_{2}=(M_{2,ij})_{\mathcal{N}\times \mathcal{N}},\quad M_{2,ij}=\frac{k_{i}^{*}\sigma_{2,-}^{*}(k_{i})\sigma_{2}(k_{j})}{k_{i}^{*}-k_{j}};
\end{align}
\end{subequations}
and ii) the AKNS-type Cauchy matrix $\mathcal{N}$-pole solution for the nonlocal dFL equation \eqref{non-eqFL} with
\begin{subequations}
\begin{align}
&\bK_{1}=\bJ[k,1],\quad\br_{1}=\bF_{k}[\rho_{1}(k)]\bar{\be}_{\mathcal{N}},\quad\ts_{1}=\bar{\be}_{\mathcal{N}}^{\st}\bH^{\st}_{k}[\sigma_{2}(k)],\\
&\bM_{1}=-\bF_{k}[\rho_{1}(k)]\bG_1\bF^{\dagger}_{k}[\rho_{1,-}(k)],\quad\bM_{2}=\bK_{1}^{\dagger}
\bH^{*}_{k}[\sigma_{2,-}(k)]\bG_2\bH^{\st}_{k}[\sigma_{2}(k)],\\
&\bG_{1}=(g^{(1)}_{i,j})_{\mathcal{N}\times \mathcal{N}},\quad g^{(1)}_{i,j}=\frac{(-1)^{i-1}\mathrm{C}^{i-1}_{i+j-2}}{(k-k^*)^{i+j-1}},\\
&\bG_{2}=(g^{(2)}_{i,j})_{\mathcal{N}\times \mathcal{N}},\quad g^{(2)}_{i,j}=\frac{(-1)^{i-1}\mathrm{C}^{i-1}_{i+j-2}}{(k^*-k)^{i+j-1}}.
\end{align}
\end{subequations}
Here $k_{i}-k_{j}^*\neq0$. The discrete plane wave factors $\rho_{1}(z)$ and $\sigma_{2}(z)$ are defined as in equation \eqref{plane-def}, and the matrix functions $\bJ,\bF,\bH$ are defined in \eqref{J-def} and \eqref{FH-def}, respectively.
\end{theorem}
	
In what follows, we examine the explicit AKNS-type Cauchy matrix one-, two-soliton and the 2-pole solution
for the nonlocal dFL equation \eqref{non-eqFL}. 
	
$\bullet$\hspace{1em}\textit{One-soliton solution.} Taking $\mathcal{N}=1$ in \eqref{AKNS-nonlocal-msr-def}
and with direct calculation, one has
\begin{align}
\label{1-soliton-AKNS-non}
\mu=\frac{4\Im(k_{1})^{2}\mathcal{A}(k_{1})}{4\Im(k_{1})^{2}+k_{1}^{*}\mathcal{A}_{-}^{*}(k_{1})\mathcal{A}(k_{1})},
\end{align}
whose envelope reads
\begin{align}
\label{1-soliton-AKNS-non-modu-1}
\abs{\mu}^{2}=\frac{\abs{\mathcal{A}(k_{1})}^{4}}{\abs{ \mathcal{A}(k_{1})+\mathcal{K}_{1}\mathcal{A}^{*}(k_{1})}^{2}},
\end{align}
where $\mathcal{K}_{1}\doteq\frac{k_{1}(\mathcal{A}^{0}(k_{1}))^{2}}{4\Im(k_{1})^{2}}$ and
relation $\mathcal{A}(k_{1})\mathcal{A}_{-}(k_{1})=(\mathcal{A}^{0}(k_{1}))^{2}$ is applied. 
	
To identify the singularity of $\abs{\mu}^{2}$, we denote the real and imaginary parts of
$\mathcal{A}(k_{1})+\mathcal{K}_{1}\mathcal{A}^{*}(k_{1})$ as 
\begin{subequations}
\label{ReIm}
\begin{align}
\label{Re}
&\Re:\quad e^{\Re(\xi_{1})}(\cos\zeta_1+\abs{\mathcal{K}_{1}}\cos(\zeta_1-\epsilon_1)),\\ 
\label{Im}
&\Im:\quad e^{\Re(\xi_{1})}(\sin\zeta_1-\abs{\mathcal{K}_{1}}\sin(\zeta_1-\epsilon_1)),
\end{align}
\end{subequations}
in which $\zeta_1=\text{Arg}(\mathcal{A}(k_{1}))=n\theta_1 + m \Theta_1+\Theta_{1}^{0}+2\kappa_{11}\pi,~
\epsilon_1=\text{arg}(\mathcal{K}_{1})+2\kappa_{12}\pi$, and $\xi_{1}$ is defined in \eqref{xi-def}, where $\kappa_{1\iota}\in\mathbb{Z}~(\iota=1,2)$ and
\begin{align}
\theta_1=\arg\frac{p+k_{1}}{p-k_{1}}, \quad \Theta_1=\arg\frac{k_{1}q+1}{k_{1}q-1},\quad\Theta_{1}^{0}=\arg\mathcal{A}^{0}(k_1).
\end{align}
	
Substituting \eqref{ReIm} into \eqref{1-soliton-AKNS-non-modu-1} and noticing $\abs{\mathcal{A}(k_{1})}=e^{\Re(\xi_{1})}$, we rewrite 
\eqref{1-soliton-AKNS-non-modu-1} as 
\begin{align}
\label{1-soliton-AKNS-non-modu-2}
\abs{\mu}^{2}=\frac{e^{2\Re(\xi_{1})}}{1+\abs{\mathcal{K}_{1}}^2+2\abs{\mathcal{K}_{1}}\cos(2\zeta_1-\epsilon_1)}.
\end{align}
We observe that when $\abs{\mathcal{K}_{1}}=1$, i.e., 
$\abs{k_{1}(\mathcal{A}^{0}(k_{1}))^{2}}=4\Im(k_{1})^{2}$, the solution \eqref{1-soliton-AKNS-non-modu-2}
has singularities along points
\begin{align}
2(n\theta_1 + m \Theta_1+\Theta_{1}^{0})=\text{arg}(\mathcal{K}_{1})+(2\kappa+1)\pi,\quad \kappa\in\mathbb{Z}. 
\end{align}
While when $\abs{\mathcal{K}_{1}}\neq1$, the solution \eqref{1-soliton-AKNS-non-modu-1} has a nonsingular oscillatory profile modulated by an exponential envelope and
satisfies the following bounds:
\begin{align}
\frac{e^{2\Re(\xi_{1})}}{(\abs{\mathcal{K}_{1}}+1)^{2}}\leq\abs{\mu}^{2}\leq
\frac{e^{2\Re(\xi_{1})}}{(\abs{\mathcal{K}_{1}}-1)^{2}},
\end{align} 
and characterized approximately as the following: 
	
\noindent{\it extreme point trajectories}: 
\[\text{arg}(\mathcal{K}_{1}) - 2 (n\theta_1 + m \Theta_1+\Theta_{1}^{0} )=\arctan\frac{\beta_{1}}{\Theta_1}
+(-1)^{\kappa}\arcsin\frac{\beta_{1}(1+\abs{\mathcal{K}_{1}^2})}{2\abs{\mathcal{K}_{1}}\sqrt{\Theta_1^2+\beta_{1}^2}}+\kappa\pi,\quad \kappa\in\mathbb{Z},\]
	
	
\noindent{\it velocity}: $-\theta_1/\Theta_1$,
	
\noindent{\it period in $n,~m$ direction}: $T_n=\pi/|\theta_1|,~T_m=\pi/|\Theta_1|,$  
	
\noindent{\it distance between two adjacent local maximum (or minimum) trajectories}: $\pi/\sqrt{\theta_1^2+\Theta_1^2}$.  
	
\vskip35pt
\begin{center}
\begin{picture}(120,80)
\put(-150,-23){\resizebox{!}{4.0cm}{\includegraphics{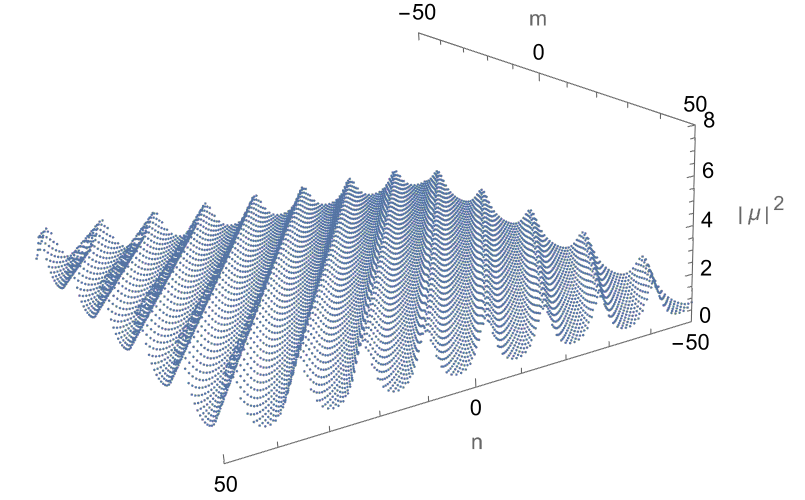}}}
\put(100,-23){\resizebox{!}{4cm}{\includegraphics{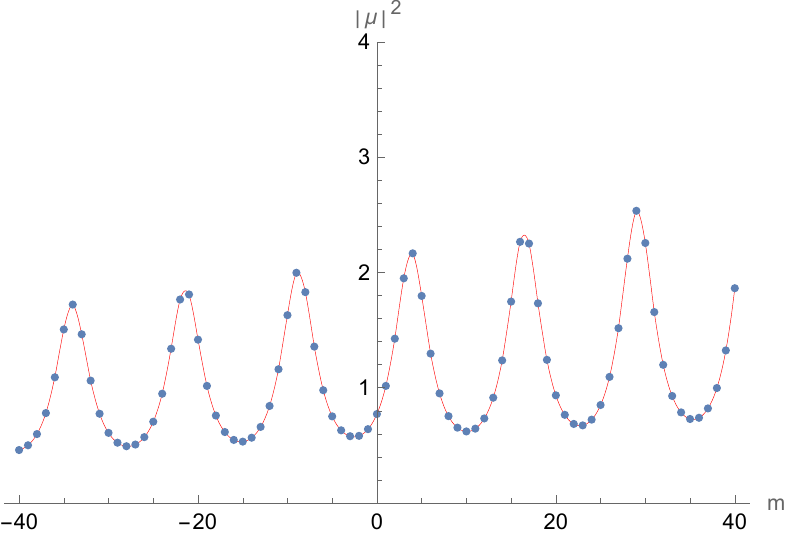}}}
\end{picture}
\end{center}
\vskip 20pt
\begin{center}
\begin{minipage}{15cm}{\footnotesize
\qquad\qquad\qquad\qquad\qquad(a)\qquad\qquad\qquad\qquad\qquad\qquad\qquad\qquad\qquad\qquad\qquad\qquad\quad (b) \\
{\bf Figure 4.} One-soliton solution \eqref{1-soliton-AKNS-non-modu-1} with $k_{1}=0.01 + 0.8\mathrm{i},~p=q=10$ and $\mathcal{A}^{0}(k_{1})=1$. 
(a) shape and movement;  (b) 2D-plot of (a) at $n=1$.}
\end{minipage}
\end{center}
	
$\bullet$\hspace{1em}\textit{Two-soliton solution}. We take $\mathcal{N}=2$ in \eqref{AKNS-nonlocal-msr-def} and get the
two-soliton solution $\mu=g/f$, in which
\begin{subequations}
\label{2-soliton-AKNS-non}
\begin{align}		
& f=1-\sum_{i=1}^{2}\sum_{j=1}^{2}\frac{k_{i}^{*}\mathcal{A}_{-}^{*}(k_{i})\mathcal{A}(k_{j})}{(k_{j}-k_{i}^{*})^{2}}
+\frac{\abs{k_{1}-k_{2}}^{4}\prod_{i=1}^{2}k_{i}^{*}\mathcal{A}_{-}^{*}(k_{i})\mathcal{A}(k_{i})}{\prod_{i=1}^{2}\prod_{j=1}^{2}(k_{i}-k_{j}^{*})^{2}},\\ 
& g=\mathcal{A}(k_{1})+\mathcal{A}(k_{2})-(k_{1}-k_{2})^{2}\mathcal{A}(k_{1})\mathcal{A}(k_{2})
\sum_{i=1}^{2}\frac{k_{i}^{*}\mathcal{A}_{-}^{*}(k_{i})}{(k_{i}-k_{i}^{*})^{2}(k_{3-i}-k_{i}^{*})^{2}}.
\end{align}
\end{subequations}

\begin{center}
\begin{picture}(120,80)
\put(-120,-23){\resizebox{!}{4.0cm}{\includegraphics{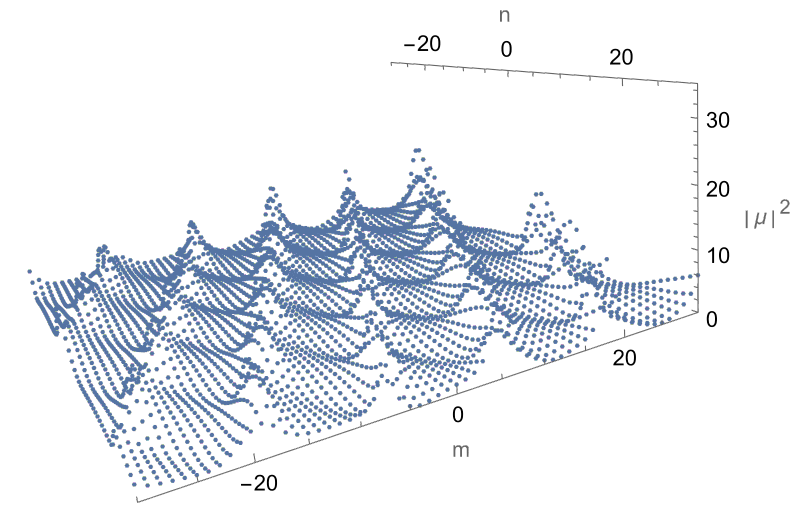}}}
\put(80,-23){\resizebox{!}{4cm}{\includegraphics{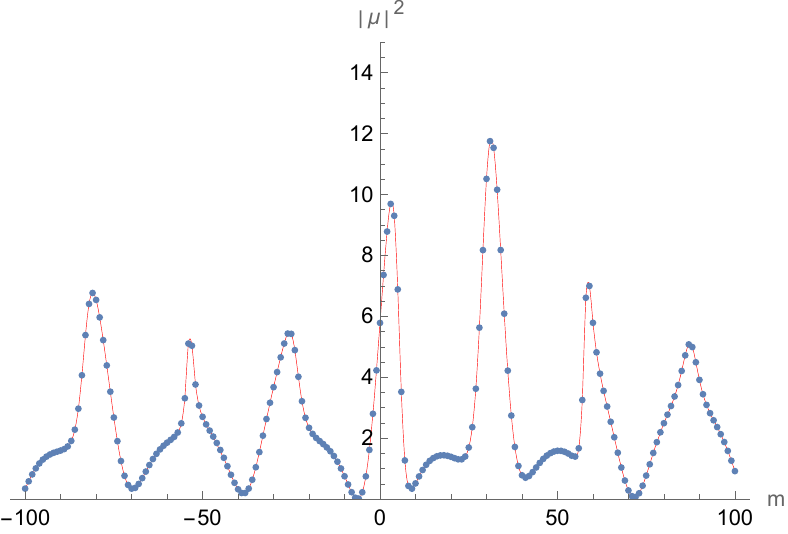}}}
\end{picture}
\end{center}
\vskip 20pt
\begin{center}
\begin{minipage}{15cm}{\footnotesize
\qquad\qquad\qquad\qquad\qquad\qquad\quad(a)\qquad\qquad\qquad\qquad\qquad\qquad\qquad\qquad\qquad\qquad (b) \\
{\bf Figure 5.} Two-soliton solution $\abs{\mu}^2$ given by \eqref{2-soliton-AKNS-non} with $k_{1}=0.008-2.28\mathrm{i},~
k_2=0.001+\mathrm{i},~p=q=10$ and $\mathcal{A}^{0}(k_{1})=\mathcal{A}^{0}(k_{2})=1$. (a) shape and movement;  (b) 2D-plot of (a) at $n=5$.}
\end{minipage}
\end{center}
	
$\bullet$\hspace{1em}\textit{2-pole solution}. We also assume that $\mathcal{A}^0(k)$ is independent of $k$; the two-pole solution is then given by $\mu=g/f$, where
\begin{subequations}
\label{2-ps-AKNS}
\begin{align}
f=&(k-k^{*})^{8}+k^{*^2}\mathcal{A}^2(k)(\mathcal{A}^*_{-}(k))^2+(k-k^{*})^{4}\mathcal{A}(k)\mathcal{A}^*_{-}(k)\big(6k^{*}+2(k-k^{*}) \nn \\ 
& \cdot\big(1-k^*(\eta+\eta^*)\big)-(k-k^{*})^{2}(\eta-k^*\abs{\eta}^2)\big),\\
g=&(k-k^{*})^{3}\mathcal{A}(k)\big(\eta(k-k^{*})^{5}+\mathcal{A}(k)\mathcal{A}_{-}^*(k)\big((k+3k^*)-\eta^*k^*(k-k^*)\big)\big). 
\end{align}
\end{subequations}
Here $\eta$ is defined in \eqref{lr-sy-2pl}.
We depict $\abs{\mu}^2$ given by \eqref{2-ps-AKNS} in Figure 6.
	
\vskip35pt
\begin{center}
\begin{picture}(120,80)
\put(-160,-23){\resizebox{!}{3.5cm}{\includegraphics{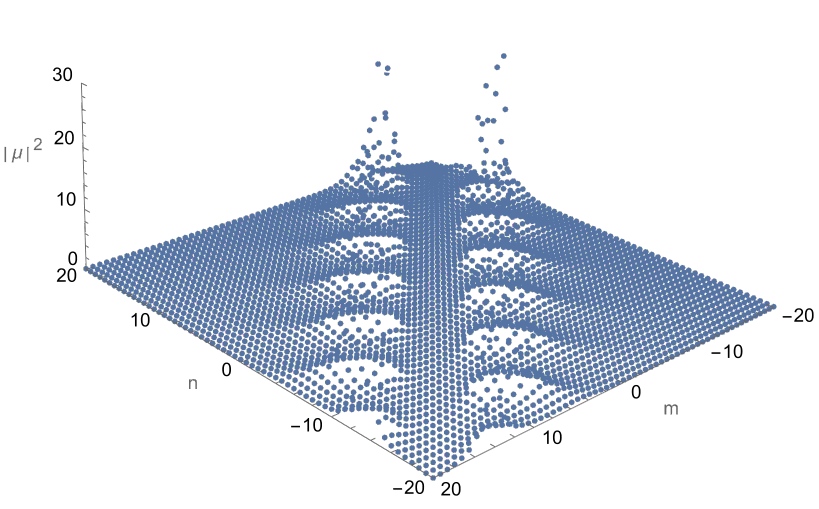}}}
\put(-10,-23){\resizebox{!}{3.5cm}{\includegraphics{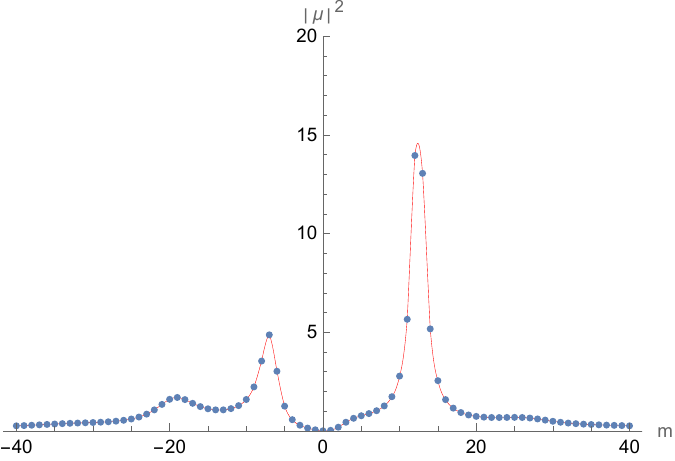}}}
\put(140,-23){\resizebox{!}{3.5cm}{\includegraphics{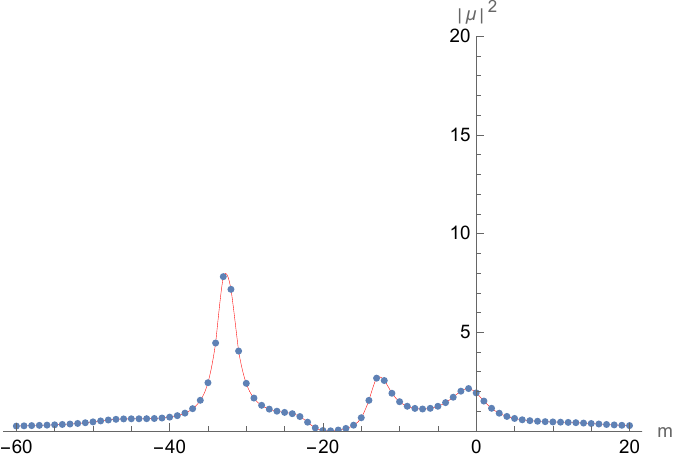}}}
\end{picture}
\end{center}
\vskip 20pt
\begin{center}
\begin{minipage}{15cm}{\footnotesize
\qquad\qquad\qquad\qquad(a)\qquad\qquad\qquad\qquad\qquad\qquad\qquad (b) \qquad\qquad\qquad\qquad\qquad\qquad\qquad\qquad (c)\\
{\bf Figure 6.} 2-pole solution $\abs{\mu}^2$ given by \eqref{2-ps-AKNS} with $k=\mathrm{i},~p=q=10$ and $\mathcal{A}^{0}(k)=1$:
(a) shape and movement; (b) 2D-plot of (a) at $n=0$; (c) 2D-plot of (a) at $n=19$.}
\end{minipage}
\end{center}
	
\subsubsection{KP-type solution}
	
The KP-type solution \eqref{asym-FL-solu} admits nonlocal reduction $\nu=\mathrm{i}\mu^{*}_{-}$, which is revealed in the following theorem.
\begin{theorem}
The KP-type solution \eqref{asym-FL-solu} admits nonlocal reduction $\nu=\mathrm{i}\mu^{*}_{-}$ under the constraints	 
\begin{align}
\label{KP-nonlocal-redu1}
\bL=\bK^{\dagger},\quad\br_{2}=\bK(\bs^{\st}_{2,-})^{\dagger},\quad\bs^{\st}_{1}=\br_{1,-}^{\dagger}.
\end{align}
\end{theorem}
\begin{proof}
By analogy with the proof employed in Theorem \ref{Th-AK-non}, we derive
\begin{align}	\bM_{1,-}^{\dagger}=-\bM_{1},\quad\bM_{2,-}^{\dagger}=-\bK^{-1}\bM_{2}\bK^{\dagger}
\quad\Longrightarrow\quad\bM_{-}^{\dagger}=-\bM_{1}-\bK^{-1}\bM_{2}\bK^{\dagger}.
\end{align}
Noting that $\bM=\bM_{1}+\bM_{2}$ and after some computations, it is shown that
\begin{align*}
u_{3,-}^{*}v_{4}=&u_{3,-}^{\dagger}v_{4}\\
=&\br_{1,-}^{\dagger}(\bM_{-}^{\dagger})^{-1}(\bs^{\st}_{2,-})^{\dagger}(1-\bs^{\st}_{2}\bM^{-1}\bK^{-1}\br_{2})\\
=&\bs^{\st}_{1}(\bM_{-}^{\dagger})^{-1}(\bM-\bK^{-1}\br_{2}\bs^{\st}_{2})\bM^{-1}\bK^{-1}\br_{2}\\
=&\bs^{\st}_{1}(\bM_{-}^{\dagger})^{-1}(\bM_{1}+\bK^{-1}\bM_{2}\bK^{\dagger})\bM^{-1}\bK^{-1}\br_{2}\\
=&-\bs^{\st}_{1}\bM^{-1}\bK^{-1}\br_{2}=v_{2}.
\end{align*}
This proof is hereby concluded.
\end{proof}
	
The KP-type Cauchy matrix soliton solution and multiple-pole solution
to the nonlocal dFL equation \eqref{non-eqFL} are summarized in the following theorem. 
\begin{theorem}\label{Theo-KP-solu-non}
The function 
\begin{align}
\label{KP-nonlocal-FLeq-solu} 
\mu=\bs^{\st}_{2}(\bM_{1}+\bM_{2})^{-1}\br_{1}
\end{align}
provides i) the KP-type Cauchy matrix $N$-soliton solution for the nonlocal dFL equation \eqref{non-eqFL} with
\begin{subequations}
\label{KP-nonlocal-msr-def}
\begin{align}
\label{KP-non-k-def}
&\bK=\mathrm{diag}(k_{1},k_{2},\cdots,k_{N}),\quad k_{i}\in\mathbb{C},\quad i=1,2,\cdots,N,\\
\label{KP-non-sr1-def}
&\bs^{\st}_{2}=\left(\frac{\rho_{2,-}^{*}(k_{1})}{k_{1}^{*}},\frac{\rho_{2,-}^{*}(k_{2})}{k_{2}^{*}},\cdots,\frac{\rho_{2,-}^{*}(k_{N})}{k_{N}^{*}}\right) ,\quad
\br_{1}=(\rho_{1}(k_{1}),\rho_{1}(k_{2}),\cdots,\rho_{1}(k_{N}))^{\st},\\
\label{KP-non-m1-def}
&\bM_{1}=(M_{1,ij})_{N\times N},\quad M_{1,ij}=\frac{\rho_{1}(k_{i})\rho_{1,-}^{*}(k_{j})}{k_{i}-k_{j}^{*}},\\
\label{KP-non-m2-def}
&\bM_{2}=(M_{2,ij})_{N\times N},\quad M_{2,ij}=\frac{\rho_{2}(k_{i})\rho_{2,-}^{*}(k_{j})}{(k_{i}-k_{j}^{*})k_{j}^{*}};
\end{align}
\end{subequations}
and ii) the KP-type Cauchy matrix $N$-pole solution for the nonlocal dFL equation \eqref{non-eqFL} with
\begin{subequations}
\begin{align}
&\bK=\bJ[k,1],\quad\br_{1}=\bF_{k}[\rho_{1}(k)]\bar{\be}_{N},\quad\bs^{\st}_{2}=\bar{\be}_{N}^{\st}\bF^{\dagger}_{k}[\rho_{2,-}(k)](\bK^{\dagger})^{-1},\\
&\bM_{1}=\bF_{k}[\rho_{1}(k)]\bar{\bG}\bF^{\dagger}_{k}[\rho_{1,-}(k)],\quad \bM_{2}=\bF_{k}[\rho_{2}(k)]\bar{\bG}\bF^{\dagger}_{k}[\rho_{2,-}(k)](\bK^{\dagger})^{-1},\\
&\bar{\bG}=(\bar{G}_{i,j})_{N\times N},\quad \bar{G}_{i,j}=\frac{(-1)^{i-1}\mathrm{C}^{i-1}_{i+j-2}}{(k-k^{*})^{i+j-1}}.
\end{align}
\end{subequations}
Here $k_{i}-k_{j}^*\neq0$. The discrete plane wave factors $\rho_{\iota}(z)$ are defined as in equation \eqref{plane-rho-def}, and the matrix functions $\bJ,~\bF,~\bH$ are defined in \eqref{J-def} and \eqref{FH-def}, respectively.
\end{theorem}
	
$\bullet$\hspace{1em}\textit{One-soliton solution}. We take $N=1$ in \eqref{KP-nonlocal-msr-def} and derive
\begin{align}
\label{12-soliton-KP-non}
\mu=\frac{2\mathrm{i}\Im(k_{1})\mathcal{B}(k_{1})}{1+k_{1}^{*}\mathcal{B}(k_{1})\mathcal{B}_{-}^{*}(k_{1})},
\end{align}
whose envelope reads 
\begin{align}
\label{1ss-kp-non}
\abs{\mu}^{2}=\frac{4\Im(k_{1})^2\abs{\mathcal{B}(k_{1})}^2}{1+\abs{\mathcal{L}}^2+2\abs{\mathcal{L}}\cos(2\zeta_{2}-\epsilon_{2})},
\end{align}
in which $\mathcal{L}=k_{1}(\mathcal{B}^{0}(k_{1}))^2$, $\zeta_{2}=\mathrm{Arg}(\mathcal{B}(k_{1}))=n\theta_{1}+m\Theta_{1}+\bar{\Theta}^{0}_{1}+2\kappa_{21}\pi$, 
$\epsilon_{2}=\arg(\mathcal{L})+2\kappa_{22}\pi$ and
\begin{align}
\label{B-def}
\mathcal{B}(z)=\frac{\rho_1(z)}{\rho_2(z)}=\left(\frac{p+z}{p-z}\right)^{n}\left( \frac{q+z^{-1}}{q-z^{-1}}\right)^{m}\mathcal{B}^{0}(z),
\end{align}
where $\bar{\Theta}_{1}^{0}=\arg\mathcal{B}^{0}(k_{1})$ and $\kappa_{2\iota}\in\mathbb{Z}~(\iota=1,2)$.
\begin{remark}
It is evident that $\mathcal{A}(z)$ defined in \eqref{A-def} differs from $\mathcal{B}(z)$ defined in \eqref{B-def} by a complex constant factor.
\end{remark}
	
With the analysis of the singularity of nonlocal one-soliton solutions of AKNS-type mentioned earlier, 
we know that when $\abs{\mathcal{L}}=1$, the solution \eqref{1ss-kp-non} has singularities along points
\begin{align}
2(n\theta_{1}+m\Theta_{1}+\bar{\Theta}^{0}_{1})=\arg(\mathcal{L})+(2\kappa+1)\pi,\quad \kappa\in\mathbb{Z}. 
\end{align}
While when $\abs{\mathcal{L}}\neq1$, the solution \eqref{1ss-kp-non} satisfies the inequality
\begin{align} \frac{4\Im(k_{1})^{2}\abs{\mathcal{B}(k_{1})}^{2}}{(\abs{\mathcal{L}}+1)^{2}}
\leq\abs{\mu}^{2}\leq\frac{4\Im(k_{1})^{2}\abs{\mathcal{B}(k_{1})}^{2}}{(\abs{\mathcal{L}}-1)^{2}},
\end{align}
and the corresponding characteristic parameters are listed as follows:
	
\noindent{\it extreme point trajectories}: 
\[\arg(\mathcal{L}) - 2 (n\theta_1 + m \Theta_1+\bar{\Theta}_{1}^{0} )=\arctan\frac{\beta_{1}}{\Theta_1}+(-1)^{\kappa}\arcsin\frac{\beta_{1}(1+\abs{\mathcal{L}^2})}{2\abs{\mathcal{L}}\sqrt{\Theta_1^2+\beta_{1}^2}}+\kappa\pi,\quad \kappa\in\mathbb{Z},\]
	
\noindent{\it velocity}: $-\theta_1/\Theta_1$,
	
\noindent{\it period in $n,~m$ direction}: $T_n=\pi/|\theta_1|,~T_m=\pi/|\Theta_1|,$  
	
\noindent{\it distance between two adjacent local maximum (or minimum) trajectories}: $\pi/\sqrt{\theta_1^2+\Theta_1^2}$.  
	
This wave is illustrated in Figure 7.
\vskip45pt
\begin{center}
\begin{picture}(120,80)
\put(-150,-23){\resizebox{!}{4.0cm}{\includegraphics{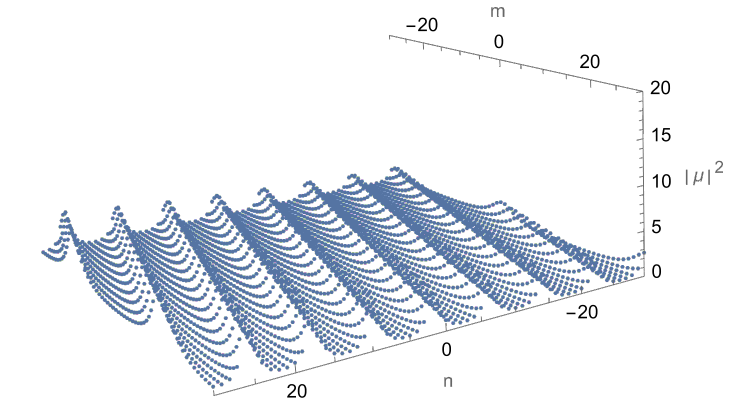}}}
\put(80,-23){\resizebox{!}{4cm}{\includegraphics{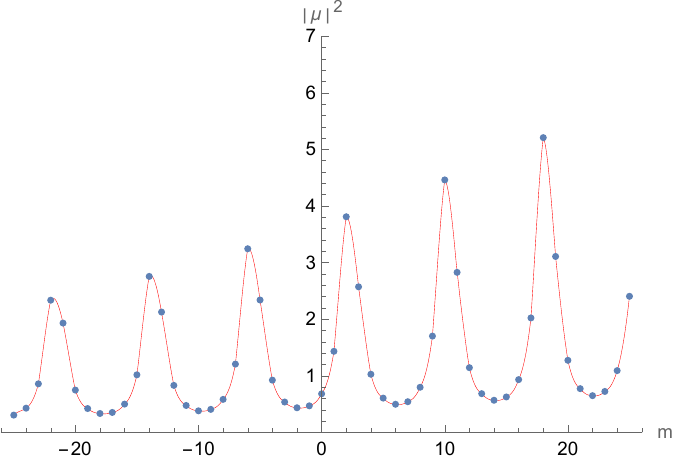}}}
\end{picture}
\end{center}
\vskip 20pt
\begin{center}
\begin{minipage}{15cm}{\footnotesize
\qquad\qquad\qquad\qquad\qquad\qquad(a)\qquad\qquad\qquad\qquad\qquad\qquad\qquad\qquad\qquad\qquad(b) \\
{\bf Figure 7.} One-soliton solution \eqref{1ss-kp-non} with $k_{1}=0.011-0.5\mathrm{i},~p=q=10$ and $\mathcal{B}^{0}(k_{1})=1$. 
(a) shape and movement;  (b) 2D-plot of (a) at $n=1$.}
\end{minipage}
\end{center}
	
$\bullet$\hspace{1em}\textit{Two-soliton solution}. The two-soliton solution arising from \eqref{KP-nonlocal-msr-def} 
with $N=2$ is expressed as $\mu=g/f$, where
\begin{subequations}
\label{2-soliton-KP-non}
\begin{align}
f=&\sum_{i=1}^{2}\sum_{j=1}^{2}\frac{(-1)^{i+j}\mathcal{B}(k_{j})\mathcal{B}_{-}^{*}(k_{i})}{k_{3-i}^{*}(k_{i}-k_{j}^{*})(k_{3-i}-k_{3-j}^{*})}+
\left(\frac{1}{(k_{1}-k_{1}^{*})(k_{2}-k_{2}^{*})}-\frac{1}{(k_{1}-k_{2}^{*})(k_{2}-k_{1}^{*})} \right) \nn \\
&\cdot \left(\frac{1}{k_{1}^{*}k_{2}^{*}}+\mathcal{B}(k_{1})\mathcal{B}(k_{2})\mathcal{B}_{-}^{*}(k_{1})\mathcal{B}_{-}^{*}(k_{2})\right), \\ 
&g=\mathcal{B}(k_{1})\mathcal{B}(k_{2})\sum_{i=1}^{2}\frac{(k_{3-i}-k_{i})\mathcal{B}_{-}^{*}(k_{i})}{(k_{i}-k_{i}^{*})
(k_{3-i}-k_{i}^{*})k_{3-i}^{*}}+\frac{1}{k_{1}^{*}k_{2}^{*}}\sum_{i=1}^{2}\frac{(k_{i}^{*}-k_{3-i}^{*})
\mathcal{B}(k_{3-i})}{(k_{i}-k_{i}^{*})(k_{i}-k_{3-i}^{*})}.
\end{align}
\end{subequations}
We depict such a wave in Figure 8.
	
\vskip45pt
\begin{center}
\begin{picture}(120,80)
\put(-120,-23){\resizebox{!}{4.0cm}{\includegraphics{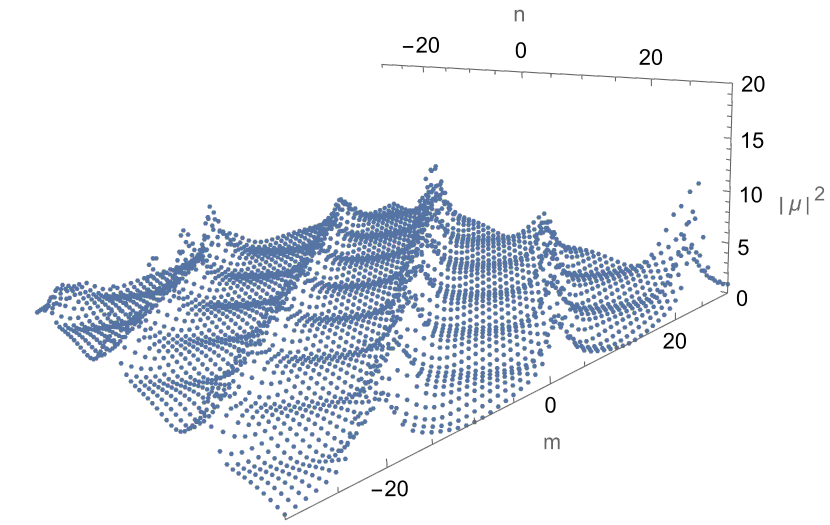}}}
\put(80,-23){\resizebox{!}{4cm}{\includegraphics{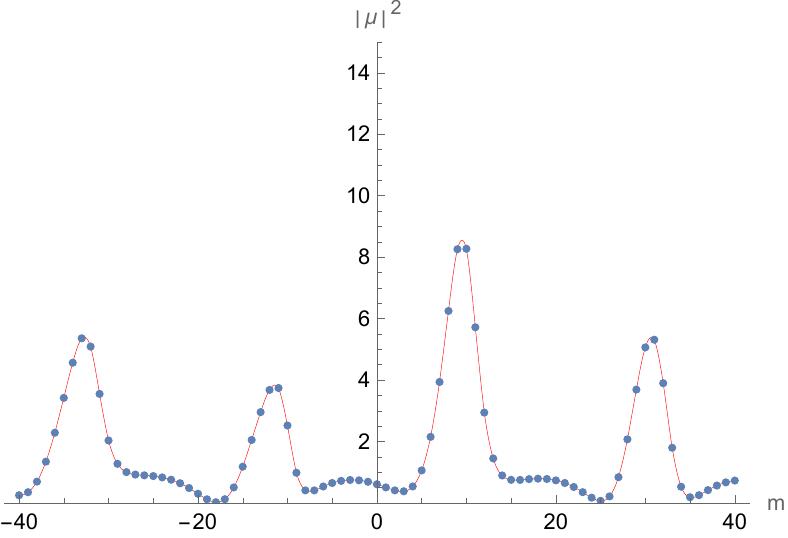}}}
\end{picture}
\end{center}
\vskip 20pt
\begin{center}
\begin{minipage}{15cm}{\footnotesize
\qquad\qquad\qquad\qquad\qquad\qquad\quad(a)\qquad\qquad\qquad\qquad\qquad\qquad\qquad\qquad\qquad\quad (b) \\
{\bf Figure 8.} Two-soliton solution $\abs{\mu}^2$ given by \eqref{2-soliton-KP-non} with $k_{1}=0.001-1.4\mathrm{i},~
k_{2}=-0.002+0.68\mathrm{i},~p=q=10$ and $\mathcal{B}^{0}(k_{1})=\mathcal{B}^{0}(k_{2})=1$. (a) shape and movement;  (b) 2D-plot of (a) at $n=1$.}
\end{minipage}
	\end{center}
	
$\bullet$\hspace{1em}\textit{2-pole solution}. Similarly, we assume that $\mathcal{B}^{0}(k)$ is independent of $k$.
Thus 2-pole solution is described as $\mu=g/f$, where
\begin{subequations}
\label{2-ps-KP-nl}
\begin{align}
&f=1+\mathcal{B}(k)\mathcal{B}^*_{-}(k)\big(k^*(2+k^*\mathcal{B}(k)\mathcal{B}^*_{-}(k))-(k-k^*)^2(\eta-k^*\abs{\eta}^2)\big), \\
&g=(k-k^*)\mathcal{B}(k)\big(2+\eta(k-k^*)+\mathcal{B}(k)\mathcal{B}^*_{-}(k)\big(k+k^*-\eta^*k^*(k-k^*)\big)\big).
\end{align}
\end{subequations}
Here $\eta$ is defined in \eqref{lr-sy-2pl}.
Such a wave is illustrated in Figure 9.

\vskip35pt
\begin{center}
\begin{picture}(120,80)
\put(-160,-23){\resizebox{!}{3.5cm}{\includegraphics{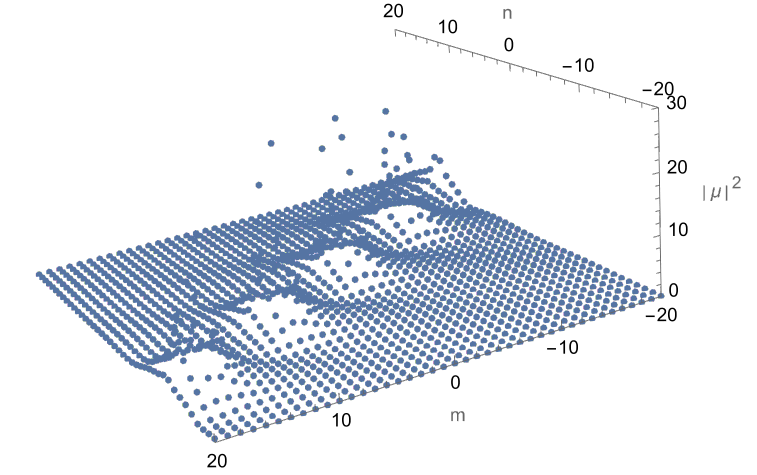}}}
\put(-10,-23){\resizebox{!}{3.5cm}{\includegraphics{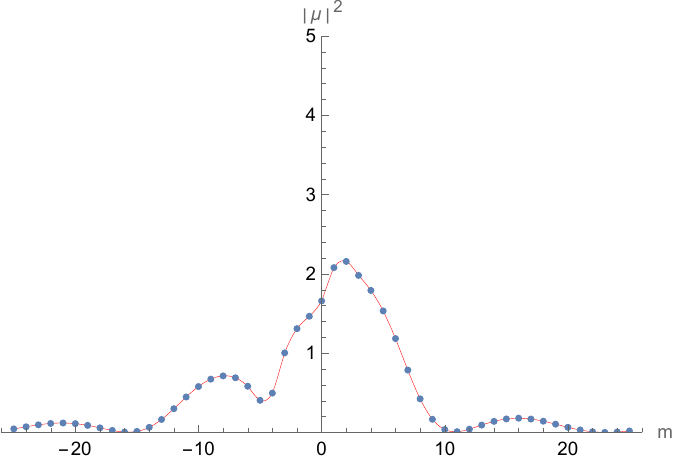}}}
\put(140,-23){\resizebox{!}{3.5cm}{\includegraphics{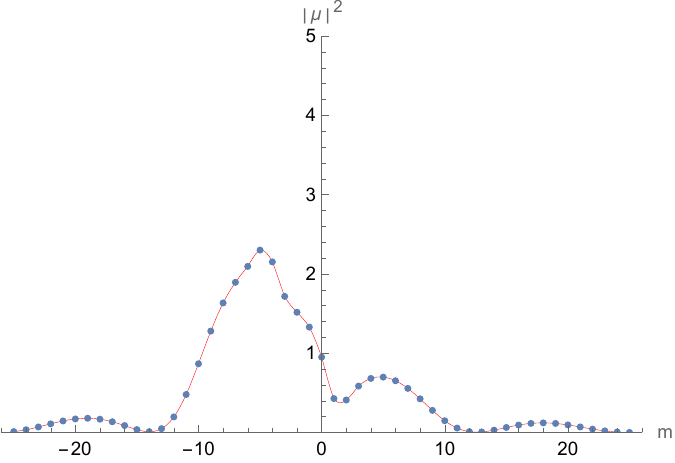}}}
\end{picture}
\end{center}
\vskip 20pt
\begin{center}
\begin{minipage}{15cm}{\footnotesize
\qquad\qquad\qquad\quad(a)\qquad\qquad\qquad\qquad\qquad\qquad\qquad(b) \qquad\qquad\qquad\qquad\qquad\qquad\qquad\quad(c)\\
{\bf Figure 9.} 2-pole solution $\abs{\mu}^2$ given by \eqref{2-ps-KP-nl} with 
$k=0.001+0.8\mathrm{i},~p=q=10$ and $\mathcal{B}^{0}(k)=1$:
(a) shape and movement; (b) 2D-plot of (a) at $n=1$; (c) 2D-plot of (a) at $n=4$.}
\end{minipage}
\end{center}

\section{Concluding remarks}
\label{sec-6}

In this paper, we have constructed four families of conjugation-symmetric discrete Mikhailov models and studied their reductions to local 
and nonlocal dFL equations. The construction relies on alternative representations of the auxiliary quantities entering the Cauchy matrix
compatibility conditions. It preserves the underlying matrix solution structure 
while yielding scalar equation pairings that are compatible with complex-conjugation reductions. Successive continuum limits connect the discrete 
Mikhailov equations first to a semi-discrete system and then to the continuous Mikhailov model, the plane-wave factors reducing accordingly. 
The distinction between the local and the nonlocal case shows up in the pairing of the component equations: every symmetric discrete Mikhailov
model admits the local reduction $\nu=\mu^{*}$, whereas the nonlocal constraint $\nu=\mathrm{i}\mu_{-}^{*}$ is imposed on suitably 
recombined systems. These constructions yield four local and four nonlocal scalar dFL equations, and clarify the role played by the 
lattice-shift symmetry in passing from coupled systems to scalar models. At the solution level, we have established consistent reductions 
of the Cauchy matrix data. The resulting formulas give local solutions within the AKNS-type scheme and nonlocal solutions within both the AKNS-type 
and the KP-type schemes. Diagonal spectral matrices produce soliton solutions, whereas Jordan-block data produce multiple-pole solutions. 
For representative local solutions, the explicit formulas fix the intensity and the trajectory of the one-soliton, and the two-soliton asymptotics 
describe elastic interactions with finite position shifts. For the nonlocal one-soliton solutions, the denominator yields explicit singularity conditions 
that involve both the spectral parameters and the lattice-dependent phase.

\vskip 20pt
\section*{Acknowledgments}
This project is supported by Natural Science Foundation of Zhejiang Province (No. LZ24A010007) and
National Natural Science Foundation of China (No. 12071432).

\vskip 20pt
\section*{Data Availability Statement}
Data sharing not applicable to this article as no datasets were generated or
analyzed during the current study.

\vskip 20pt
\section*{Conflict of interest}
There are no conflicts of interest to declare.

\appendix
		
\section{Two additional forms of discrete Mikhailov models}

In this appendix, we present two additional discrete Mikhailov models different from the four families derived in Section 2.2. 
Both forms are obtained by algebraic rearrangement of the compatibility conditions \eqref{u3u1} and \eqref{vu1}, and feature all four shift boundary terms $P,Q,\wh{P},\wt{Q}$.

Starting from equations \eqref{u3u1} and \eqref{vu1}, we have the following equivalent identities:
\begin{subequations}
\label{App-u3vu1-1}
\begin{align}
& (\wh{u}_{3}-u_{3})(p+u_{1}-\wt{u}_{1})+(u_{3}+\wt{u}_{3})(1/q+u_{1}-\wh{u}_{1}) \nn \\
& \qquad +(\wt{u}_{3}-\wh{\wt{u}}_{3})(p+\wh{u}_{1}-\wh{\wt{u}}_{1})+(\wh{u}_{3}+\wh{\wt{u}}_{3})(1/q+\wt{u}_{1}-\wh{\wt{u}}_{1})=0, \\
& (\wh{\nu}-\nu)(p+u_{1}-\wt{u}_{1}+\mathrm{i}u_{3}\nu-\mathrm{i}\wt{u}_{3}\wt{\nu})+(\nu+\wt{\nu})(1/q+u_{1}
-\wh{u}_{1}+\mathrm{i}u_{3}\nu-\mathrm{i}\wh{u}_{3}\wh{\nu}) \nn \\
& \qquad +(\wt{\nu}-\wh{\wt{\nu}})(p+\wh{u}_{1}-\wh{\wt{u}}_{1}+\mathrm{i}\wh{u}_{3}\wh{\nu}-\mathrm{i}\wh{\wt{u}}_{3}\wh{\wt{\nu}})
+(\wh{\nu}+\wh{\wt{\nu}})(1/q+\wt{u}_{1}-\wh{\wt{u}}_{1}+\mathrm{i}\wt{u}_{3}\wt{\nu}-\mathrm{i}\wh{\wt{u}}_{3}\wh{\wt{\nu}})=0.
\end{align}
\end{subequations}
Eliminating all $u_{1}$-related terms in system (A.1) by using \eqref{A1} and \eqref{B1}, we arrive at the first supplementary discrete model for $(\mu,\nu):=(u_{3},\nu)$:
\begin{subequations}
\label{dFL5}
\begin{align}
\label{dFL5-1}
& (\wh{\nu}+\wt{\nu})(\wh{\wt{\mu}}+\mu)(\wh{\wt{\mu}}+\wh{\mu}-\wt{\mu}-\mu)
-\mathrm{i}\big((\wh{\mu}-\mu)P+(\wt{\mu}+\mu)Q+(\wt{\mu}-\wh{\wt{\mu}})\wh{P}+(\wh{\mu}+\wh{\wt{\mu}})\wt{Q}\big)=0, \\
\label{dFL5-2}
& (\wh{\mu}+\wt{\mu})(\wh{\wt{\nu}}+\nu)(\wh{\wt{\nu}}+\wh{\nu}-\wt{\nu}-\nu)
+\mathrm{i}\big((\wh{\nu}-\nu)P+(\wt{\nu}+\nu)Q+(\wt{\nu}-\wh{\wt{\nu}})\wh{P}+(\wh{\nu}+\wh{\wt{\nu}})\wt{Q}\big)=0,
\end{align}
\end{subequations}
where $P$ and $Q$ are defined as in \eqref{PQ}.

Equations \eqref{u3u1} and \eqref{vu1} also admit another set of equivalent identities:
\begin{subequations}
\label{App-u3vu1-2}
\begin{align}
& (\wt{u}_{3}-u_{3})(p+u_{1}-\wt{u}_{1})+(u_{3}+\wh{u}_{3})(1/q+u_{1}-\wh{u}_{1}) \nn \\
& \qquad +(\wh{u}_{3}-\wh{\wt{u}}_{3})(p+\wh{u}_{1}-\wh{\wt{u}}_{1})+(\wt{u}_{3}+\wh{\wt{u}}_{3})(1/q+\wt{u}_{1}-\wh{\wt{u}}_{1})=0, \\
& (\wt{\nu}-\nu)(p+u_{1}-\wt{u}_{1}+\mathrm{i}u_{3}\nu-\mathrm{i}\wt{u}_{3}\wt{\nu})+(\nu+\wh{\nu})(1/q+u_{1}-\wh{u}_{1}+\mathrm{i}u_{3}\nu-\mathrm{i}\wh{u}_{3}\wh{\nu}) \nn \\
& \qquad +(\wh{\nu}-\wh{\wt{\nu}})(p+\wh{u}_{1}-\wh{\wt{u}}_{1}+\mathrm{i}\wh{u}_{3}\wh{\nu}-\mathrm{i}\wh{\wt{u}}_{3}\wh{\wt{\nu}})
 +(\wt{\nu}+\wh{\wt{\nu}})(1/q+\wt{u}_{1}-\wh{\wt{u}}_{1}+\mathrm{i}\wt{u}_{3}\wt{\nu}-\mathrm{i}\wh{\wt{u}}_{3}\wh{\wt{\nu}})=0.
\end{align}
\end{subequations}
Similarly, eliminating all terms related to $u_{1}$ yields the second supplementary discrete model for $(\mu,\nu):=(u_{3},\nu)$:
\begin{subequations}
\label{dFL6}
\begin{align}
\label{dFL6-1}
& (\wt{\nu}+\wh{\nu})(\mu^{2}-\wh{\wt{\mu}}^{2})+(\nu+\wh{\wt{\nu}})(\wt{\mu}^{2}-\wh{\mu}^{2})
+\mathrm{i}\big((\wt{\mu}-\mu)P+(\wh{\mu}+\mu)Q+(\wh{\mu}-\wh{\wt{\mu}})\wh{P}+(\wt{\mu}+\wh{\wt{\mu}})\wt{Q}\big)=0, \\
\label{dFL6-2}
& (\wt{\mu}+\wh{\mu})(\nu^{2}-\wh{\wt{\nu}}^{2})+(\mu+\wh{\wt{\mu}})(\wt{\nu}^{2}-\wh{\nu}^{2})
-\mathrm{i}\big((\wt{\nu}-\nu)P+(\wh{\nu}+\nu)Q+(\wh{\nu}-\wh{\wt{\nu}})\wh{P}+(\wt{\nu}+\wh{\wt{\nu}})\wt{Q}\big)=0.
	\end{align}
\end{subequations}

The two models \eqref{dFL5} and \eqref{dFL6} differ from the four families of symmetric discrete Mikhailov models \eqref{dFL1}-\eqref{dFL4} in the main text. 
The most notable distinction is that both supplementary models contain all four terms $P,Q,\wh{P},\wt{Q}$, while each family in the main text omits one of these terms.

The asymmetric discrete Mikhailov model presented in our previous work \cite{ZMZ2026} is essentially a cross-pairing of the above two supplementary models: 
it combines \eqref{dFL5-1} with \eqref{dFL6-2} (or equivalently \eqref{dFL5-2} with \eqref{dFL6-1}), which leads to the lack of conjugate symmetry. 

\section{Remaining dFL equations}
	
The following three dFL equations are obtained by applying the local reduction $\nu=\mu^*$
to models \eqref{dFL2}, \eqref{dFL3} and \eqref{dFL4}, respectively:
	\begin{subequations}
	\begin{align}
	&(\wh{\mu}+\wt{\mu})\big((\mu^*+\wh{\wt{\mu}}^*)(\wh{\mu}-\mu)-\mathrm{i}\wh{P}_{1}-\mathrm{i}Q_{1}\big)
	+(\mu+\wh{\wt{\mu}})\big((\wh{\mu}^*+\wt{\mu}^*)(\wh{\wt{\mu}}-\wt{\mu})-\mathrm{i}\wt{Q}_{1}+\mathrm{i}\wh{P}_{1}\big)=0,\\
	&(\wh{\mu}+\wt{\mu})\big((\mu^*+\wh{\wt{\mu}}^*)(\wh{\wt{\mu}}-\wt{\mu})-\mathrm{i}P_{1}-\mathrm{i}\wt{Q}_{1}\big)
	+(\mu+\wh{\wt{\mu}})\big((\wh{\mu}^*+\wt{\mu}^*)(\wh{\mu}-\mu)-\mathrm{i}Q_{1}+\mathrm{i}P_{1}\big)=0,\\
	&(\wh{\mu}+\wt{\mu})\big((\mu^*+\wh{\wt{\mu}}^*)(\mu+\wt{\mu})+\mathrm{i}P_{1}+\mathrm{i}\wt{Q}_{1}\big)
	-(\mu+\wh{\wt{\mu}})\big((\wh{\mu}^*+\wt{\mu}^*)(\wh{\mu}+\wh{\wt{\mu}})-\mathrm{i}\wt{Q}_{1}+\mathrm{i}\wh{P}_{1}\big)=0,
	\end{align}
	\end{subequations}
where $P_{1}$ and $Q_{1}$ are defined in \eqref{P1Q1}.

\section{Remaining nonlocal dFL equations}

The following three nonlocal dFL equations are obtained from models $(N.2)$, $(N.3)$ and $(N.4)$ under the nonlocal reduction $\nu=\mathrm{i}\mu_{-}^*$:
\begin{subequations}
\begin{align}
& (\wh{\mu}+\wt{\mu})\big((\mu_{-}^*+\wh{\wt{\mu}}_{-}^*)(\wh{\mu}-\mu)-\wh{P}_{2}-Q_{2}\big)
+(\mu+\wh{\wt{\mu}})\big((\wh{\mu}_{-}^*+\wt{\mu}_{-}^*)(\wh{\wt{\mu}}-\wt{\mu})-\wt{Q}_{2}+\wh{P}_{2}\big)=0, \\
& (\wh{\mu}+\wt{\mu})\big((\mu_{-}^*+\wh{\wt{\mu}}_{-}^*)(\wh{\wt{\mu}}-\wt{\mu})-P_{2}-\wt{Q}_{2}\big)+
(\mu+\wh{\wt{\mu}})\big((\wh{\mu}_{-}^*+\wt{\mu}_{-}^*)(\wh{\mu}-\mu)-Q_{2}+P_{2}\big)=0, \\
& (\wh{\mu}+\wt{\mu})\big((\mu_{-}^*+\wh{\wt{\mu}}_{-}^*)(\mu+\wt{\mu})+P_{2}+\wt{Q}_{2}\big)
-(\mu+\wh{\wt{\mu}})\big((\wh{\mu}_{-}^*+\wt{\mu}_{-}^*)(\wh{\mu}+\wh{\wt{\mu}})-\wt{Q}_{2}+\wh{P}_{2}\big)=0,
\end{align}
\end{subequations}
where $P_{2}$ and $Q_{2}$ are defined in \eqref{P2Q2}.

\end{document}